\documentclass{article}
\usepackage{amsfonts}
\usepackage{amssymb}
\usepackage{amsmath}

\newtheorem{theorem}{Theorem}
\newtheorem{acknowledgement}[theorem]{Acknowledgement}

\input{tcilatex}
\begin{document}

\title{Beyond gauge theory: Hilbert space positivity and causal localization
in the presence of vector mesons \\
{\small Dedicated to the memory of Raymond Stora and John Roberts}}
\author{Bert Schroer \\
present address: CBPF, Rua Dr. Xavier Sigaud 150, \\
22290-180 Rio de Janeiro, Brazil\\
permanent address: Institut f\"{u}r Theoretische Physik\\
FU-Berlin, Arnimallee 14, 14195 Berlin, Germany}
\date{November 2015}
\maketitle

\begin{abstract}
The Hilbert space formulation of interacting $s=1$ vector-potentials stands
in an interesting contrast with the point-local Krein space setting.of gauge
theory. Already in the absence of interactions the Wilson loop in a Hilbert
space setting has a "topological property" which is missing in the gauge
theoretic description (Haag duality, Aharonov-Bohm effect); the conceptual
differences increase in the presence of interactions.

The Hilbert space positivity weakens the causal localization properties if
interacting fields which results in the replacement of the gauge-variant
point-local matter fields in Krein space by string-local physical fields in
Hilbert space. The gauge invariance of the perturbative S-matrix corresponds
to its independence of the spacelike string direction of its
interpolating.fields. In contrast to gauge theory, whose immediate physical
range is limited to gauge invariant perturbative S-matrix and local
observables, its Hilbert space string-local counterpart in is a full-fledged
quantum field theory (QFT).

The new setting reveals that the Lie-structure of self-coupled vector mesons
results from perturbative implementation of the causal localization
principles of QFT.
\end{abstract}

\section{Introduction}

It is well known that the use of point-local massless vector potentials is
incompatible with the positivity of Hilbert space. One usually resolves this
problem by abandoning positivity and maintaining the point-local field
formalism which leads to gauge theory (GT). The prize to pay is well-known
from quantum electrodynamics (QED) in the standard indefinite metric
(Gupta-Bleuler) gauge setting: positivity can be recovered for local
observables, whereas charge-carrying fields remain outside its physical
range. The separation between physical and unphysical quantum fields is done
in terms of gauge symmetry which is not a physical symmetry but a formal
device to extract a physical subtheory.

Although the standard gauge formalism is restricted to vector potentials,
the clash between zero mass point-local fields and positivity is a general
phenomenon for all $s\geq 1~$zero mass tensor potentials. It does not affect
the corresponding field strengths, but the short distance dimension of the
latter ($d_{sd}=2$ for $s=1)~$would prevent their use in renormalizable
interactions. For semi-integer spin the borderline is $s\geq 3/2;$ such
fields will play no role in this paper.

Another possibility is to keep the Hilbert space, but allow a weaker
localization. The tightest covariant localization of fields beyond
point-like is causal localization on semi-infinite spacelike "strings"%
\footnote{%
Beware that strings in String Theory are not string-local in the sense of
local quantum physics.} $x+\mathbb{R}_{+}e,~e^{2}=-1.~$It~is easy to
construct $m>0~$covariant fields $\Psi (x,e)~$in terms of semi-infinite line
integrals. Whereas the point-local (pl) Proca vector-potentials have no
massless limit, the correlation functions of their string-local (sl)
siblings pass without problems to their massless counterpart. The
construction of free massive sl potentials in the same Hilbert space as
their point-local counterparts guaranties that both belong to the same
relative localization class. They represent two different "field
coordinatizations" of the same theory, which (in the presence of
interactions) implies that their particle content and S-matrix are identical.

The perturbative use of string-local elementary matter fields $\Psi (x,e)$
in Hilbert space quantum field theory\footnote{%
Unless stated otherwise, QFT refers to the Hilbert space setting of quantum
theory. $\ $Gauge theory, which contains a Hilbert space subtheory, will be
referred to as GT.} (QFT) correspond to formally gauge invariant
string-local composites in terms of pl fields in GT%
\begin{equation}
\psi ^{K}(x)e^{ig\int_{0}^{\infty }A_{\mu }^{K}(x+\lambda e)e^{\mu }d\lambda
}  \label{com}
\end{equation}%
where the superscript $K~$refers to the pl indefinite metric (Krein space
instead of Hilbert space) setting of GT. Proposals to recover
gauge-invariant charge-carrying matter fields in terms of such formal
expressions appeared already a long time before renormalization theory \cite%
{Jor}. They played no role in the discovery of renormalized perturbation
theory, but they re-appeared later in Mandelstam's proposal to replace the
perturbative gauge theoretic setting of QED by one which uses solely Hilbert
space gauge-invariant field strengths instead of Krein space vector
potentials \cite{Mand}.

Steinmann's confronted the difficult task of incorporating non-linear and
non-local composites as (\ref{com}) into the perturbative gauge theoretic
setting \cite{Stein}. Morchio and Strocchi's (still ongoing) project was
motivated by the task to extract long distance (infrared) information by
studying global limits in a positivity-restoring topology \cite{M-S}.

As already indicated, the present paper uses a new perturbation theory based
on \textit{massive} string-local vector potentials which are defined as
semi-infinite spacelike line-integrals over field strengths. These covariant
fields maintain the Hilbert space positivity and permit a smooth $%
m\rightarrow 0$ limits of their vacuum expectation values. These fields are
accompanied by new objects referred to as "escort" fields; they are
stringlocal $s=1~$scalar fields (not possible with pl field) which first
appeared in the construction of string-local intertwiners \cite{MSY} as
unexpected objects of curiosity, and years later reappeared when in
discussions with Jens Mund it became clear that they playd a natural role in
a linear relation between pl Proca fields and their sl siblings.

It turned out that these scalar escorts play an indispensable role in a
Hilbert space setting of perturbation theory which only uses physical
degrees of freedom. Far from being restricted to kinematical aspects of sl
fields, the escorts play a deep role in the the dynamics; in some models
they even appear already in the first order interaction density. Instead of
criticizing previous gauge theoretic attempts we leave it to the reader to
notice the shortcomings of the gauge theoretic formalism which instead
enlarges the degrees of freedom by negative metric Stueckelberg fields and
ghosts which, since they must be removed in the end, play a role which is
reminiscent of chemical catalyzers.

In the new setting the lowest order perturbative interaction densities are
defined in terms of Wick products of string-local covariant free vector
potentials $A_{\mu }(x,e)$ (obtained by integrating point-local field
strengths along spacelike semi-lines) with point-local $s<1$ free matter
fields. The higher order interactions spread the string-localization of the
potentials to the $s<1\ $matter fields and in this way force them to be sl
without having to impose sl as in (\ref{com}); interestingly their massless
limits become much stronger sl than the $A_{\mu }$ which retain the linear
relation with pl field strengths. These sl matter fields are the elementary
interacting matter fields of the new which reolace the composite Krein
fields (\ref{com}). Whereas the sl vector potentials maintain their $d_{sd}~$%
(ignoring the expected perturbative logarithmic corrections), the short
distance dimensions with which the pl fields enter the interaction ($%
d_{sd}=1 $ for $s=0$ and $d_{sd}=3/2$ for $s=1/2$) in a Hilbert space
setting would increase linearly with the order of perturbation; constructing
them as sl fieldss again maintain their original $d_{sd}$ modulo logarithms.

The problem is then to extend renormalization theory from pl to sl fields.
The feasibility of such a project relies on the observation that sl fields
have a better short distance behavior than their pl counterparts. Whereas $%
d_{sd~}$of pl free field tensor-potentials increases with spin as $%
d_{sd}^{s}=s+1,$ that of their string-local siblings remains at $%
d_{sd}^{s}=1 $ \textit{independent of} $s.~$This implies that there is no
problems to construct interaction densities \textit{for any spin within the
power-counting bound (PCB)} $d_{sd}^{int}\leq 4$ of renormalizability$.~$In
the present paper the spin of the $s\geq 1~$higher spin fields will be
integer; for spinor fields there are corresponding results.

However in contrast to interactions between $s<1~$pl fields, for which the
first order PCB is the only restriction, there are additional requirements
for sl fields which has no pl counterpart and in fact bears no relation to
short distance properties. \textit{Higher order terms will in general not
remain sl but rather lead to a total delocalization}; such situations turn
out to be inconsistent with the principles of QFT. Heuristically this new
phenomenon can be understood in terms of the $x$-integration in \textit{inner%
} propagator lines which, in case of $x$ being the starting point of a
semi-infinite spacelike line $x+\mathbb{R}_{+}e,$ causes a complete higher
order delocalization.

Its avoidance imposes a severe conditions on the interaction density. In
contrast to the PCB short distance renormalizability requirement, the new
restriction maintains the sl of those fields whose vacuum expectation values
one wants to computes but prevents the dependence on "inner" $e^{\prime }s$
which, after inner integrations over $x,$ would spread the innter strings
over Minkowski spacetime.

It vaguely corresponds to the requirement of gauge invariance. The latter
bears no direct relation to the foundational principles of QFT (The pl
localization of gauge-dependent fields in a Krein space is not the physical
localization !), but the prevention of total delocalization is directly
related to the causal localization principles of QFT. Whereas the asymptotic
short distance property ("asymptotic freedom") is expected to agree with
that in the sl Hilbert space setting\footnote{%
This would require the Callan-Symanzik equation and in particular its beta
function in the sl setting to be $e$-independent...}, the understanding of
long distance properties requires the use of sl fields.

Some of these nonrenormalizable pl interactions turn out to be
renormalizable in terms of $s\geq 1~$covariant sl fields. This is in
particular the case if one allows higher order short distance compensations
between couplings containing fields of different spins (reminiscent of
compensations between different spin components in supersymmetry
multiplets). Those which are not renormalizable in the sl sense are in all
likelihood not models of QFT. This points to an interesting connection
between perturbative renormalizability and causal localization properties.
The main purpose of this paper is to illustrate these new ideas in their
simplest possible low order perturbative context. Technically more demanding
tasks, as the elaboration of an $n^{th}$ order Epstein-Glaser
renormalization theory which includes causal string-crossings, will be left
to future publications.

Quantum gauge theory also achieves a reduction from $d_{sd}=2$ to $\dot{d}%
_{sd}=1$ in case of $s=1,$ but it accomplishes this not with a Hilbert space
positivity maintaining weakening of localization from pl to sl, but rather
with the help of indefinite metric which permits compensations in
intermediate states between positive and negative contributions by brute
force. If it were not be for the existence of a subset of gauge invariant
operators (which includes the S-matrix), an indefinite metric setting would
remain without physical content.

The Hilbert space setting of QT is the basis of its probabilistic
interpretation. It has no counterpart in the classical theory and hence one
cannot rely on "quantization" to solve conceptual problems of interactions
involving $s=1~$fields. Quantum gauge theory is the result of keeping pl $%
s\geq 1$ fields at the prize of sacrificing positivity. Gauge symmetry is a
local symmetry in classical electromagnetism; it bears no relation to the
principles of QFT but rather enters QFT through the "quantization"
parallelism to classical field theory\footnote{%
There are however classical limits of pl or sl free quantum fields in terms
of expectation values in coherent states.}. Perturbative QFT needs no
reference to Lagrangian quantization; it can be defined in terms of
interaction densities formed from covariant pl or sl free fields which can
be directly obtained from Wigner's positive energy representation theory of
the Poincar\'{e} group.

Perturbative gauge invariance is a way to extend the pl formalism to $s=1$
interactions in a Krein space in order to extract those quantities which do
not "feel" the presence of negative metric states (namely the local
observables and the S-matrix). It has been "an amazingly successful
placeholder" (Stora) for an unknown Hilbert space formulation within the
standard setting of perturbative QFT. The new sl Hilbert space setting
extends the Hilbert space Wightman formulation for interacting $s<1$ pl
fields to $s\geq 1~$sl fields for which the test-function smearing amounts
to smearing in $x,$ using the smooth rapidly deceasing test functions $f(x))$
\textit{and} smearing in $e,$ using smooth compact supported functions $h(e)$
in the $d=1+2$ de Sitter space of spacelike directions $e,~e^{2}=-1$%
\footnote{%
The present experience with perturbation theory suggests that interacting
Wightman fields with $s\geq 1$ exist only in the form of sl fields.}$.$ The
main purpose of the present work is to convince the reader that the
beginnings of a theory for which gauge theory is the placeholder already
exist.

The idea of constructing string-localized fields in Hilbert space can be
traced back to the solution of the localization properties of Wigner's zero
mass infinite spin representation class \cite{BGL} \cite{MSY} in terms of
the \textit{modular localization theory} of algebraic QFT (a
historic/philosophical view can be found in \cite{SHPMP}). But, as it is
often the case, the historic context which led to a new view of QFT may turn
out to be less important for its use in actual problems; the construction of
string-local fields and their use in a new perturbative setting does not
require knowledge about modular localization.

A prior result from algebraic QFT (AQFT) which highlighted the \textit{%
naturalness of string-localization} was obtained by Buchholz and Fredenhagen
in \cite{Bu-Fr}. Their theorem was formulated and proven in the
operator-algebraic setting; re-expressing its physical content in terms of
quantum fields it states:

\textbf{Theorem: }\textit{The field theoretic content of an asymptotically
complete QFT with a mass gap and local observables can be fully accounted
for in terms of operators localized on spacelike cones } \textit{whose
generating fields are expected to be\ string-local covariant fields}$~\Psi
(x,e)$ \textit{localized on} \textit{semi-lines}$~x+R_{+}e~e^{2}=-1$\ 
\textit{(the cores of narrow spacelike cones).} \textit{Point-local
observable fields are }$e$\textit{-independent }$\Psi ^{\prime }s$. \textit{%
QFT does not need generating fields with weaker than stringlike localization
properties (as e.g. localization on spacelike hypersurfaces)}.

The Hilbert space of such a theory can be described as a Wigner-Fock Hilbert
space spanned by those particle states which the (LSZ or Haag-Ruelle)
scattering theory associates asymptotically with the string-local fields.
Since functional analytic and operator-algebraic tools are not available in
indefinite metric Krein spaces, the gauge theoretic setting is strictly
limited to the combinatorial rules of finite order renormalized perturbation
theory whereas there are no known nonperturbative tools to derive theorems
(TCP, Spin\&Statistics, theorems about scattering theory,...) which e.g.
rely on the use of the Schwartz inequality. On the other hand most
mathematical tools of (Wightman) QFT\ remain valid for string-local fields;
hence large parts of textbook presentation of nonperturbative QFT can be
adjusted to the new situation.

More specific results come from the new perturbation theory of sl fields.
Whereas positivity-preserving renormalizable interactions between
point-local fields are limited to low spins $s<1,$ renormalizable
interactions containing $s\geq 1$ fields in a Hilbert space setting require
a weakening of localization from point- to string-local fields \cite{Rio} 
\cite{vector}. Couplings which even remain nonrenormalizable in a
string-local setting are not expected to represent models of QFT.

The a priory knowledge that the Hilbert space in models with a mass gap can
be identified with a Wigner-Fock particle space turns out to be very useful
in the perturbative construction\footnote{%
It is important that this Fock spaces refers to asymptotic fields
(particles); the use of a Fock space at finite times is forbidden by "Haag's
theorem" \cite{Haag}.}. This well understood situation is also a good
starting point for investigating massless limits in which the standard
Wigner particle structure is lost (which manifests itself in terms of
infrared divergencies) and the appropriate Hilbert space has to be
constructed from the massless limit of the massive correlation functions 
\cite{St-Wi}. Most deep unsolved problems of QFT, in particular a spacetime
understanding of infrared properties which account for the conceptual
changes of large time scattering properties in QED and are responsible for
confinement in QCD, are connected with interactions involving massless
vector potentials. Already for free fields the massless field is most
conveniently constructed through the massless limit of the free massive sl
correlation functions.

In view of the strong relation between causal localization and Hilbert space
positivity one cannot expect that GT accounts for the correct causal
localization properties. The absence of of positivity (unitarity) is not
only affecting issues of Hermitian adjoints and unitarity, but also spoils
the foundational causality property of QFT. The positivity and the ensuing
string-localization is particularly important for large distance properties
which usually announce their presence through infrared divergencies. Long
distance problems do not occur in $s<1$ interactions; their physical
understanding is inexorably related to properties of $s\geq 1$ interacting
sl fields. Although not yet established, one expects that the short distance
asymptotic freedom behavior of the gauge dependent matter fields will be
confirmed in the sl Hilbert space theory.

The role of sl fields will be illustrated in two problems. One problem, the
so-called "violation of Haag duality", is closely related to the
Aharonov-Bohm effect. It shows that the use of Hilbert space string-local
vector potentials in a Wilson loop leads to a topological effect which is
absent for gauge theoretic point-like potentials in Krein space and has
generalizations to all massless $s\geq 1$ fields. The somewhat "eery
feelings" about an apparent causality violation (which partially account for
the popularity of the A-B effect) can be traced back to the use of the
point-local potentials of gauge theory; they disappear once one uses instead
their covariant string-local counter-part which perceive the difference
between Haag duality and the more general Einstein causality.

The main problem posed by string-local fields is the construction of the
correct first order interaction densities. It is well-known that the
requirement of PCB $d_{sd}^{int}\leq 4~$(the prerequisite for
renormalizability) only permits pl couplings between fields of low spin $s<1$
with $d_{sd}=1$ or $3/2~$(for $s=1/2$).~String-local fields, whose short
distance dimension is $d_{sd}=1$ for bosons (and $3/2$ for fermions \textit{%
independent of }$s),$ permit PCB couplings for any spin.

But, as already mentioned before, there exists an additional rather severe
restriction on sl interaction-densities which has no counterpart for
renormalizable interactions between $s<1$ pl fields. Most string-local
interactions within $d_{sd}^{int}\leq 4~$lead to a \textit{total
delocalization in higher orders}, which removes them from the list of
potential candidates for a QFT model. This restriction will be exemplified
for three models: couplings of massive string-local vector mesons to complex
fields (massive QED), to Hermitian fields (the Higgs model) and self-coupled
massive vector mesons.

The presence of self-coupled massive vector mesons lead to a new phenomenon;
it is the only known coupling for which the first order PCB does not imply
renormalizability; or to formulate it in a historical context: passing from
the 4-Fermi interaction to intermediate self-interacting massive vector
mesons improves the short distance properties so that the coupling leads to
first order PCB, but fails in second order. The culprit is a second order
"induced" $d_{sd}=5$ self-coupling which can only be compensated by
enlarging the first order coupling by a $A\cdot AH$ coupling with a scalar
Hermitian $H$ field whose second order iteration contains a compensating $H$%
-independent term. Whereas the massive (scalar, spinor) vector mesons of QED
do not require the presence of a (or several) $H,~$the second order PCB of
string-local massive Y-M and QCD cannot be maintained without a compensating 
$H$-coupling$.~$

The combination of a shift in the complex field space of a massless scalar
QED (SSB "breaking of gauge symmetry") and a subsequent reality adjustment
by an operator gauge transformation (no counterpart in SSB) leads also to
the correct interaction density, but this "Higgs mechanism" bears no
physical relation with the intrinsic origin of the second order induced (not
postulated !) Mexican hat potential of a $H$ self-interaction (section 4).
A\ \ genuine physical SSB breaking on the other hand leaves its intrinsic
(i.e. independent of by what prescription it was obtained) physical mark on
the final model and there is simply none. The identically conserved Maxwell
current $j_{\mu }=\partial ^{\nu }F_{\mu \nu }~$of an abelian massive vector
meson leads to a screened (and not to spontaneously broken) charge \cite%
{vector}. Interactions between Hermitian fields and massive vector mesons
are (as their complex counterparts) fully accounted for by the BRST gauge
invariance of the S-matrix $\mathfrak{s}S=0~$or its spacetime Hilbert space
counterpart namely the independence of $S$ on string directions $d_{e}S=0.$

The perturbative string-local field theory (SLFT) shares many \textit{formal}
similarities with gauge theory (GT). This is particularly evident if one
formulates GT not in terms of Feynman rules but instead uses the so-called
causal gauge invariant operator setting (CGI) \cite{Scharf}. The reason is
that gauge invariance must hold on-shell but is violated off-shell. This
makes it difficult to discuss properties as positivity (unitarity), whose
validity in gauge theories is \textit{limited to the on-shell restrictions}
of correlation functions (i.e. the S-matrix). In the CGI operator
formulation of the BRST gauge formalism \cite{Scharf} \cite{Aste} \cite{BDSV}
the perturbative implementation is focussed on the construction on the
S-operator which fulfills $\mathfrak{s}S=0$ where $\mathfrak{s}$ is the
nilpotent BRST s-operation; in this way the construction of the
gauge-invariant S-operator is separated from the construction of the less
interesting gauge dependent correlation functions.

The implementation of the string-independence of the S-matrix and the
independence of "inner" strings of correlation functions of string-local
fields is achieved with the help of a differential form calculus on the $%
d=1+2~$de Sitter space of string directions. Unlike GT with its
cohomological BRST formalism based on a nilpotent $\mathfrak{s}$-operation,
whose physical range is restricted to local observables and the perturbative
S-matrix, the SLFT is a full QFT which contains the important string-local
physical matter fields whose LSZ limits connect the causal localization
principles implemented in terms of fields and their charge-carrying local
equivalence classes with the observable world of individual particles just
as in point-local Hilbert space QFTs. But the appearance of the $e$%
-dependence in going from on- to off-shell complicates the analytic
connection between scattering amplitudes and correlation functions.

It is interesting to use some hindsight from this new developments for
looking back at the origin of modern GT. In the 't Hooft-Veltman paper on
the renormalization of nonabelian gauge theory the important issue was the
verification of the unitarity of the S-matrix (ignoring infrared problems).
What they achieved with the help of lengthy "by hand" calculations was
afterwards "streamlined" by adding formal tools. Starting with
Faddeev-Popov, being enriched by Slavnov, the formal polishing finally
reached its present perfection in the hands of Becchi, Rouet, Stora and
Tuytin. It would be impossible for scholars of QFT to derive the BRST gauge
formalism from the principles of QFT; they remain useful inventions for
extracting a physical subtheory (in particular the S-matrix) from an
unphysical description in Krein space.

The SLFT Hilbert space setting contains only physical degrees of freedom.
But this conceptual economy leads to new and even somewhat surprising
concepts. The massive pl Proca field $A_{\mu }^{P}(x)$ turns out to lead to
a \textit{pair} of sl fields, a sl vector potential $A_{\mu }(x,e)~$\textit{%
and} an sl scalar field $\phi (x,e)$. Both fields appear in the interaction,
but in the zero mass limit only the sl $A_{\mu }~$survives$.$This has a
surprising analogy with the appearance of short range vector potentials in
the BCS theory of superconductivity. In this analogy the bosonic $\phi $
field corresponds to the Cooper pairs. As the $\phi $ field, the Cooper
pairs result from a reorganization of existing degrees of freedom.

Neither in the many-body quantum mechanical description of BCS
superconductivity which leads to London's short range vector potentials, nor
for the description of QFT interactions of massive abelian vector mesons
with matter one needs new (e.g. Higgs) degrees of freedom. The QFT
counterpart of the Cooper pair regrouping are the $\phi $ fields; it will be
shown that they are an indispensable epiphenomenon of interacting massive
vector mesons in a Hilbert space. This clears the way for asking the
question why one needs $H$-fields in the presence of self-interacting
massive vector mesons. The answer has already been given in \cite{vector}
and will receive additional attention in section 4.

Whereas the SLFT Hilbert space formalism has many formal similarities with
the implementation of on-shell BRST gauge invariance (section 5) in the
operator formulation of "causal gauge invariance" (CGI) of the University of
Z\"{u}rich group published during the 90s \cite{Scharf} \cite{Aste}, its
conceptual and calculational power reaches beyond when it comes to long
distance infrared problems which cannot be described in terms of an
S-matrix. The correct analog of the long-range Coulomb interaction of
quantum mechanics are the long distance properties of physical matter fields
interacting with massless sl vector potentials. In such cases the structure
of a Wigner-Fock Hilbert space and an S-matrix acting in it breaks down and
the remaining objects from which the theory has to be reconstructed are the
massless limits of the sl correlation functions.

The next section presents the kinematics of SLFT i.e. relations between the
string-local free fields and their use in interaction densities for which
one does not need perturbative calculations. The third section addresses
what the title of this paper promises. Some second order perturbative
results and their interpretation are presented in section 4; this section
also contains educated guesses about physical manifestations of infrared
properties. Section 5 presents formal analogies with the CGI operator
formulation of the BRST formalism. The outlook contains comments about
ongoing calculations and conjectures about what one hopes to accomplish in
the future.

Many of the ideas arose in extensive discussions over several years with
Jens Mund, but the responsibility for possible errors and less than perfect
presentations rest on my shoulders.

\section{A Hilbert space alternative to local gauge theory, kinematical
aspects}

Massive free spin $s\geq 1$ fields are commonly described in terms of degree 
$s$ tensor potentials. For $s=1$ this reduces to the well-known Proca
potential 
\begin{align}
& A_{\mu }^{P}(x)=\frac{1}{(2\pi )^{3/2}}\int
(e^{ipx}\sum_{s_{3}=-1}^{1}u_{\mu }(p,s_{3})a^{\ast }(p,s_{3})+h.c.)\frac{%
d^{3}p}{2p_{0}} \\
& \left\langle A_{\mu }^{P}(x)A_{\nu }^{P}(x^{\prime })\right\rangle =\frac{1%
}{(2\pi )^{3}}\int e^{-ip\xi }M_{\mu \nu }(p)\frac{d^{3}p}{2p_{0}},\text{ }%
M_{\mu \nu }=-g_{\mu \nu }+\frac{p_{\mu }p_{\nu }}{m^{2}}  \notag
\end{align}%
~To

To this point-local field one may associate two string-local fields, a
vector potential $A_{\mu }(x,e)$ and a scalar field $\phi (x,e)~$\cite{MSY},
as well as two "field-valued differential forms" in $e$-space namely a
one-form $u(x,e)$ and a two-form $\hat{u}(x,e)$ 
\begin{align}
A_{\mu }(x,e)& =\int_{0}^{\infty }F_{\mu \nu }(x+\lambda e)e^{\nu }d\lambda
,~with~F_{\mu \nu }(x)=\partial _{\mu }A_{\nu }^{P}(x)-\partial _{\nu
}A_{\mu }^{P}(x),  \label{lamda} \\
~~\phi (x,e)& =\int_{0}^{\infty }A_{\mu }^{P}(x+\lambda e)e^{\mu }d\lambda
,~~e^{2}=-1  \notag \\
u& =d_{e}\phi =\partial _{e^{\mu }}\phi de^{\mu },~\hat{u}=d_{e}(A_{\alpha
}de^{\alpha })  \notag
\end{align}%
which are all members of the equivalence class of relatively string-local
fields which are associated to the point-local $A_{\mu }^{P}~$Proca field
(the sl "Borchers class"). It is well known \cite{St-Wi} that in the
presence of a mass gap interacting point-local fields within the same
localization class lead to the same particle physics (particles, S-matrix);
this continues to be valid for string-local fields \cite{Bu-Fr} \cite{Haag}.

Throughout this paper the differential form calculus on the $d=1+2$ de
Sitter space of string-directions will play an important role. As the $x,$
the $e^{\prime }s$ are variables in which the fields fluctuate; they bear no
relation with the "mute" gauge-fixing parameters of GT.

String-local vector potentials fulfill $e^{\mu }A_{\mu }(x,e)=0,$ and in the
massless limit also $\partial ^{\mu }A_{\mu }(x,e)=0.$ These relations are
not imposed gauge conditions but rather intrinsic properties of string-local
potentials which result from the above definitions. Note that $A_{\mu },u$
and $\hat{u}$ possess zero mass limits, whereas $A_{\mu }^{P}$ and $\phi $
remain finite only in the combination $A^{P}-\partial \phi .$ They are all
free fields, but their mutual dependence leads to mixed 2-point functions
(below).

One may change the string "density" $d\lambda \rightarrow q(\lambda
)d\lambda ,~q(\infty )=1~$within the linear field class. There is no
conceptual problem with this continuous enlargement since quantum fields, in
contrast to classical fields, have no observable "individuality". The latter
property is an attribute of particles which share their superselected
charges with their associated field-classes with which they are (large-time)
asymptotically related \cite{Haag}. Renormalizability requires to define PCB
interaction densities in terms of maximally fourth degree polynomials of 
\textit{fields with the lowest short distance dimension within a field class}
but the distinction between elementary and composite is not based on the
short distance dimensions of fields but pertains to the fusion properties of
superselected charges of field classes.

In order to obtain linear relations between these fields as%
\begin{equation}
A_{\mu }=A_{\mu }^{P}+\partial _{\mu }\phi  \label{lin}
\end{equation}%
one must use the same $q(\lambda )~$and in order to maintain simplicity of
two-point function 
\begin{eqnarray}
&&\left\langle A_{\mu }(x,e)A_{\mu ^{\prime }}(x^{\prime },e^{\prime
})\right\rangle =\frac{1}{\left( 2\pi \right) ^{3}}\int e^{-ip(x-x^{\prime
})}M_{\mu \mu ^{\prime }}^{A,A}(p;e,e^{\prime })\frac{d^{3}p}{2p_{0}}
\label{2-point} \\
&&M_{\mu \mu ^{\prime }}^{A,A}(p;e,e^{\prime })=-g_{\mu \mu ^{\prime }}-%
\frac{p_{\mu }p_{\mu ^{\prime }}(e\cdot e^{\prime })}{(p\cdot e-i\varepsilon
)(p\cdot e^{\prime }+i\varepsilon )}+\frac{p_{\mu }e_{\mu ^{\prime }}}{%
(p\cdot e-i\varepsilon )}+\frac{p_{\mu }e_{\mu ^{\prime }}^{\prime }}{%
(p\cdot e^{\prime }+i\varepsilon )}  \notag \\
M^{\phi \phi } &=&\frac{1}{m^{2}}-\frac{e\cdot e^{\prime }}{(p\cdot
e-i\varepsilon )(p\cdot e^{\prime }+i\varepsilon )},~M_{\mu }^{A\phi }=\frac{%
1}{i}(\frac{e_{\mu }^{\prime }}{p\cdot e^{\prime }+i\varepsilon }-\frac{%
p_{\mu }e\cdot e^{\prime }}{(p\cdot e-i\varepsilon )(p\cdot e^{\prime
}+i\varepsilon )})  \notag
\end{eqnarray}%
These expressions may be either directly derived from the above line
integral or be obtained from the~$A,A^{P}$ and $\phi $-intertwiners\footnote{%
A systematic and detailed account of the construction of string-local fields
and their use in a string-extended Epstein-Glaser construction of
time-ordered products will be contained in a forthcomong manuscript by Jens
Mund..} (the $e$-dependent denominators including the $\varepsilon $%
-prescription result from the Fourier transformation of the Heavyside
function). Clearly the $\lambda $-integration (\ref{lamda}) has lowered the
short distance dimension of the vector potential from $d_{sd}^{P}=2$ to $%
d_{sd}=1$ and the heuristic reading of (\ref{lin}) is that the derivative $%
\partial \phi $ of the $d_{sd}=1~\phi $-escort removes the most singular
part of $A^{P}$ at the price of directional $e-$fluctuation. In the massless
limit the $A^{P}$ as well as $\phi $ diverge, but the string-local potential 
$A_{\mu }(x,e)~$remains well-defined; its 4-dimensional curl is the
point-local field strength.

The necessity to work with string-local potentials and the \textit{%
appearance of mixed two point functions} is the prize for working in the 
\textit{Hilbert space of physical degrees of freedom.} In contrast the
indefinite metric Krein space of gauge theory contains in addition to the
Gupta-Bleuler and St\"{u}ckelberg indefinite metric degrees of freedom also
contains those of "ghost" fields. There are no mixed contributions since
these unphysical free fields are independent. The BRST gauge formalism
permits to extract physical data but falls short of extracting a full QFT.

The sl $A_{\mu }(x,s)~$and its scalar escort $\phi (x,e)$ are different
fields, but they do not enlarge the degrees of freedom; in fact they are
linear in the same Wigner creation and annihilation operators $%
a^{\#}(p,s_{3})$ of the point-local Proca potential$.$ Although they do not
change the degrees of freedom, their separate appearance in interaction
densities plays an important role in upholding string localization of higher
order interacting fields. The QFT of the textbooks, in which each particle
corresponds to one pl field, is limited to interactions between $s<1$
fields. QFT of $s\geq 1$ requires the use of sl fields and each such field
is accompanied by $s~$lower spin sl escorts (see later).

In order to prevent confusion, the present work uses the terminology "QFT"
only for genuine quantum theories in Hilbert space. Gauge theories in Krein
space are not QTs as they stand (since the positivity is not less important
than $\hslash $), but the BRST formalism permits to extract quantum
subtheories (local observables acting in the Hilbert space vacuum sector,
the S-matrix acting in a Wigner-Fock particle space). The nonperturbative
mathematical tools which lead to the famous theorems of local quantum
physics (TCP, Spin\&Statistics,...) require the presence of positivity
(unitarity).

The reader may encounter many new concepts, but he can be assured that they
are not ideosyncratic inventions of the author but rather result from
reconciling higher spin interactions with the causal localization principles
in the Hilbert space setting of QT. In particular the existence of massless
vector- (more general $s>1~$tensor-) potentials in Hilbert space is tied to
the existence of the $m\rightarrow 0~$limit of sl massive correlation
functions. These massless sl higher spin fields are the counterparts of the
pl massless $s<1$ fields; their use is indispensable for the understanding
of the spacetime physics behind logarithmic infrared divergencies.

The above relation (\ref{lin}) resembles a gauge transformation. It should
be maintained in the presence of interactions with a complex matter field.
Whereas the interaction density is defined in terms of a free pl matter
field $\psi _{0}(x),$ the interaction with sl potentials will convert these
fields into an interacting sl fields $\psi (x,e).$ One also expects that
these sl field has a very singular interacting pl sibling $\psi ^{P~}$(the
analog of the pl fields in the nonrenormalizable pl Hilbert space setting)
and that both are related (in the sense of normal products) as\footnote{%
I am indebted to Jens Mund for informing me that this relation has been
checked in lowest nontrivial order.}%
\begin{equation}
\psi (x,e)=N\psi ^{P}(x)e^{ig\phi (x,e)}  \label{psi}
\end{equation}%
This together with (\ref{lin}) is certainly reminiscent of a gauge
transformation of a matter field interacting with a vector potential. But in
the present context it represents a relation between two "field
coordinatizations" of the same theory, one being a string-local Wightman
field (bounded $d_{sd}$) and the other a field with unbounded short distance
dimension as expected in nonrenormalizable couplings \cite{Ba-Sc}. This is a
class of fields which are too singular (unbounded $d_{sb}$) in order to be
compactly localizable in the sense of Wightman \cite{St-Wi}; such fields
have been studied by Jaffe \cite{Jaffe} who illustrated his more singular
fields in terms of Wick-ordered exponentials of free fields $\exp \varphi $.

After this brief excursion into uncharted territory, this section returns to
kinematic aspext of sl free fields.

The formal similarity of the directional variable $e~$of string-local fields
with a (noncovariant) "axial" gauge parameter should not hide the fact that
its gauge theoretic interpretation\footnote{%
The gauge theoretic $e$ is considered to be "mute" i.e. it is the same for
all gauge fields and remains unaffected by Lorentz transformations.} caused
unsolvable short- and long- distance problems, which finally led to its
abandonment. The reason behind this failure is that fluctuations in the
d=1+2 the unit Sitter space of spacelike directions in individual
string-local fields cannot be reconciled with a gauge interpretation;
fortunately \textit{what was a curse in the use as an axial gauge turns out
to be a blessing in the Hilbert space setting} of $s=1$ interactions.

This construction permits a generalization to any integer spin. Massive free
fields of spin $s$ and short distance dimension $d_{sd}^{s}=s+1$ are
conveniently described in terms of symmetric point-local potentials $A_{\mu
_{1}..\mu _{s}}^{P}$ of tensor degree $s.$ A corresponding string-local
tensor with $d_{sd}^{s}=1~$can be obtained in analogy to the vector
potential with the help of repeated semi-infinite line integrals (see
appendix) 
\begin{equation}
\phi _{\mu _{1},..\mu _{k}}(x.e)=\int_{0}^{\infty }..\int_{0}^{\infty
}d\lambda _{k+1}..d\lambda _{s}e^{\mu _{k+1}}..e^{\mu _{s}}A_{\mu _{1},\mu
_{2},..,\mu _{k},..,\mu _{s}}^{P}(x+\lambda _{k+1}e+..\lambda _{s}e)
\end{equation}%
For the sake of simplicity of notation we specialize to $s=2~$in which case
the corresponding relation to (\ref{lin}) is%
\begin{equation}
g_{\mu _{1}\mu _{2}}(x,e)=g_{\mu _{1}\mu _{2}}^{P}(x)+sym~\partial _{\mu
_{1}}\phi _{\mu _{2}}(x,e)+\partial _{\mu _{1}}\partial _{\mu _{2}}\phi (x,e)
\label{g}
\end{equation}%
where our notation pays tribute to the fact that the metric tensor of
general relativity is the principle physical candidate for a symmetric
second degree tensor. Note that in this case the string-local field has 2
string-local escorts, a scalar $\phi (x,e)$ and a vector $\phi _{\mu }(x,e).$
By construction the symmetric tensor fulfills $e^{\mu _{1}}g_{\mu _{1}\mu
_{2}}=0$

The field strength associated to the symmetric $g_{\mu \nu }$tensor field of 
$g_{\mu \nu }$ $s=2$ field of degree 4, which has the same mixed
symmetric-antisymmetric permutation symmetry as the Riemann tensor of
relativity (the "linearized Riemann tensor") is 
\begin{equation}
R_{\mu \nu \kappa \lambda }(x)=\frac{1}{2}antisym~\partial _{\mu }\partial
_{\kappa }g_{v\lambda }  \label{R}
\end{equation}%
This $R$-tensor, which is the $s=2$ analog of the the $s=1$ field strength $%
F_{\mu \nu }$, is the lowest degree point-local massless $s=2~$field.

The extension of (\ref{lin}) and (\ref{g}) to spin $s>2~$should be clear: a
massive point-local degree $s$ tensor potential corresponds to a
string-local potential of the same tensor degree and "$\phi $ escorts" of
lower tensor degree of which only the string-local degree $s$ potential has
a massless limit. The lowest degree point-local tensor field which permits a
massless limit is a field strength of degree $2s$ and short distance
dimension $d_{sd}=2s.$ In analogy to (\ref{R}), it results from the
application of $s$ derivatives and subsequent anti-symmetrization.

The intertwiner $u(p,s)$ for massive point-local tensor fields of degree $s$
and short distance dimension $d_{sd}=s+1~$relate the $2s+1$ spin space with
the space of symmetric covariant tensor. They are divergence free (as $%
A_{\mu }^{P}$)~and traceless. It is simpler to calculate their momentum
space two point functions which consists of linear combinations of tensors
of degree $2s$ formed from the Minkowski spacetime $g_{\mu \nu }$ and
products of $p_{\mu }$ made dimensionless by multiplication with appropriate
inverse mass powers. The requirement of vanishing trace and divergence
determines the two-point function up to a numerical factor.

The differential geometric structure of the $d=1+2~$directional de Sitter
space impart these string-local fields and related field valued differential
forms with a rich differential geometric structure which plays an important
role in the new positivity-maintaining SLF perturbation theory. In this
setting all fields are physical; the gauge invariant local observables
correspond to point-local (generally composite) fields, whereas the
interacting matter fields $\Psi (x,e)~$are string-local generalizations of
Wightman fields (polynomially bounded in momentum space or equivalently $%
d_{sd}<\infty $).

A surprising collateral kinematic result of this observation is the (easy to
verify) statement that the angular averaging over $e$ within a spacelike
plane leads to the Coulomb (or radiation) vector potential. It is well-known
that it acts in a Hilbert space but that its lack of covariance makes it
unsuitable for renormalized perturbation theory. The fact that it results
from directional averaging of covariant string-local potential (which plays
the central role in the new covariant SLFT Hilbert space setting) may come
as a surprise to some.

Before taking up the issue of interactions, it is interesting to compare the
SLFT setting with the BRST gauge formalism. The latter is based on a the
action of a nilpotent $\mathfrak{s}$-operation on indefinite metric fields
in a Krein space ($K=$Krein) extended by "ghost operators". In the notation
of \cite{Scharf} it reads

\begin{eqnarray}
&&\mathfrak{s}A_{\mu }^{K}=\partial _{\mu }u^{K},~\mathfrak{s}\phi
^{K}=u^{K},~\mathfrak{s}\hat{u}^{K}=-(\partial A^{K}+m^{2}\phi ^{K})
\label{Krein} \\
&&\mathfrak{s}B:=i[Q,B],~Q=\int d^{3}x(\partial ^{\nu }A_{\nu
}^{K}+m^{2}\phi ^{K})\overleftrightarrow{\partial }_{0}u^{K}  \notag
\end{eqnarray}%
$Q$ is the so-called ghost charge (associated to a conserved ghost current)
whose properties ensure the nilpotency ($\mathfrak{s}^{2}=0$) of the BRST $%
\mathfrak{s}$-operation. The $A_{\mu }^{K}$ is a point-like massive vector
meson in the Feynman gauge and $\phi ^{K}~$is a free scalar field whose
Krein space two-point function has the opposite sign (a kind of negative
metric scalar St\"{u}ckelberg field). The "ghosts" $u,\hat{u}$ are free
"scalar fermions" whose presence is necessary in order to recover the
perturbative positivity of local observables and a unitary S-matrix.

The $\mathfrak{s}$-invariance of the scattering $S$-operator in gauge theory
and the $d_{e}$-independence in the string-local setting are both related to
cohomology; but whereas the former has no relation to spacetime, the $d_{e}$
acts on the space-like string directions of the in $e~$independently
fluctuating$~$fields.

Since there are string-local fields with $d_{sd}=1$ for all spins, there
also exist string-local interaction densities within the power-counting
limit $d_{sd}^{int}=4.~$But as already mentioned in the introduction, there
is another physical restriction which has no counterpart in the point-local
case: couplings of string-local fields are only physical if their higher
order extensions preserve string-localization and if there remains a
subalgebra of pointlike generated local observables.

This requirement, which plays no role within the point-local renormalization
formalism, severely restricts interactions involving sl fields so that some
of the advantage of low short distance dimension is lost. In the following
this will be illustrated in three examples involving massive vector mesons
(all fields are free fields). The first two models describe a massive
vectormeson which couples either with a complex scalar field $\varphi ~$%
(scalar "massive QED") or with a Hermitian scalar field $H$ 
\begin{eqnarray}
L^{P} &=&gA_{\mu }^{P}j^{\mu },~~j_{\mu }=\varphi ^{\ast }%
\overleftrightarrow{\partial }_{\mu }\varphi :  \label{mass} \\
L^{P} &=&gmA_{\mu }^{P}A^{P,\mu }H  \label{AH}
\end{eqnarray}%
where the mass factor $m$ accounts for the mass dimension ("engineering
dimension") of the interaction density. In both cases $A_{\mu }^{P}$ is the
point-local $d_{sd}=2$ Proca potential so that the point-local interaction
density $L^{P}$ violates the PCB restriction of renormalizability.

Using the relation between the Proca potential and its string-local
counterpart $A_{\mu }~$including its escort field $\phi $ (\ref{lin})$,$ one
may rewrite the nonrenormalizable point-local interactions (\ref{mass}) into
a $d_{sd}^{int}\leq 4$ string-local expression plus the divergence of
another operator $V_{\mu }.$%
\begin{eqnarray}
&&L^{P}=L-\partial ^{\mu }V_{\mu },~~,  \label{scalar} \\
~ &&L=gA_{\mu }j^{\mu },\ V_{\mu }=j_{\mu }\phi ,~~  \notag
\end{eqnarray}%
The second line presents the wanted pair $L,V_{\mu }$ for massive scalar
QED, for both operators $d_{sd}=4.$ The renormalization-preventing
point-local interaction density $d_{sd}(L^{P})=5$ has been separated into
two string-local contributions in such a way that the renormalizability
spoiling $d_{sd}=5$ contribution has been collected into the divergence of $%
V_{\mu }.$ In massive QFTs such divergence terms may be disposed of in the
adiabatic limit so that the first order S-matrix of the power-counting
violating $L^{P}$ is the same as that of its better behaved string-local
counterpart~$L$. Although far from obvious, this idea of disposing
renormalizability-violating terms at infinity can be generalized to higher
orders. It is not limited to the S-matrix, but also leads to the
construction of polynomial bounded correlation function of string-local
quantum fields (private communication by Jens Mund). In other words the sl
perturbation theory complies with the localization properties of
string-local Wightman fields.

There is widespread belief that for perturbative renormalization theory one
needs (either canonical or functional integral) Lagrangian quantization. But
this is not correct; even for pl perturbation theory one only needs a scalar
interaction density $L^{P}$ in terms of free fields. These free fields need
not be Euler-Lagrange fields; rather any free field obtained from
covariantization of Wigner's pure quantum unitary representation theory of
the Poincar\'{e} can be used. In fact Lagrangians for most higher spin
fields are not known and sl fields are never Euler-Lagrange. The St\"{u}%
ckelberg-Bogoliubov-Epstein-Glaser perturbation theory is based on the
causal iteration of the first order scalar interaction density $\NEG{L}$
made from local Wick-products of free fields. There are no infinities, but
the iteration leads to a growing number of new parameters (counterterm
parameters) whose number only remains finite in case of the PCB $%
d_{sd}(L)\leq 4.$

The $L,V_{\mu }$ pair for the $H$ coupling is less simple\footnote{%
The perturbative calculations are simpler if one replaces only as many $%
A^{P} $ by $A$ as needed to obtain $d_{sd}(L)=4.$} since now also $L$
depends on $\phi $ 
\begin{equation}
L=gm(A\cdot A^{P}H+\phi A\cdot \partial H-\frac{m_{H}^{2}}{2}\phi
^{2}H),~V^{\mu }=gm(\phi A_{\mu }^{P}H+\frac{1}{2}\phi ^{2}\partial ^{\mu }H)
\label{H}
\end{equation}%
In this case the verification of the identity (\ref{scalar}) requires the
use of the free field equation for $H~$with mass $m_{H}$ ($m~$=mass of
vector meson)$.~$

A similar $L,\partial V$ pair exists for self-interacting massive vector
mesons, e.g.%
\begin{eqnarray}
L^{P} &=&\sum \varepsilon _{abc}F_{a}^{\mu \nu }A_{b,\mu }^{P}A_{c,\nu
}^{P}=L-\partial V  \label{self} \\
L &=&\sum \varepsilon _{abc}\left\{ F_{a}^{\mu \nu }A_{b,\mu }A_{c,\nu
}+m^{2}A_{a,\mu }^{P}A_{b}^{\mu }\phi ^{c}\right\}  \notag \\
V_{\mu } &=&\sum \varepsilon _{abc}F_{a}^{\mu \nu }\left\{ A_{b,\nu
}+A_{b,\nu }^{P}\right\} \phi _{c}  \notag
\end{eqnarray}%
For the verification one again one has to use the field equation which in
this case reads $\partial ^{\nu }F_{\mu \nu }=m^{2}A_{\mu }^{P}.$

For the extension to higher orders it is helpful to express the point-local
nature in terms of the differential form calculus on de Sitter space%
\begin{equation}
d_{e}(L-\partial V)=0  \label{p}
\end{equation}%
The existence of such pairs with $L$ within $d_{sd}^{int}\leq 1$ turn out to
be the prerequisite for the existence of local observables within a
string-local setting. They also prevent the higher order spread of
localization over all of spacetime. It should however be emphasized that the
construction of $L,V_{\mu }~$pairs is a problem which can be pursued
independent of $L^{P}.$ Whether a collection of free fields permits a
(maximally quadrilinear) coupling $L~$which can be completed to a $L,V_{\mu
} $ pair is a well-defined mathematical problem within the setting of
differential forms on $d=1+2~$de Sitter space. Whereas $L$ must stay within
PCB, the $d_{sd\text{ }}$of $V_{\mu }$ may have contributions above $%
d_{sd}=4;$ as long as these contributions do not lead to higher order short
distance contributions to $L$ beyond $d_{sd}=4$ the model remains sl
renormalizable.

The exactness of the zero form $L-\partial V~$in (\ref{p}) is a rather
restrictive localization requirement. Such pairs within the power-counting
bound for $L~$turn out to be unique (if they exist) modulo additive changes
of $V_{\mu }~$terms with vanishing divergence. Since in massive models the
divergence $\partial V~$disappears in the adiabatic on-shell limit, the
first order contribution to the S-matrix are equal 
\begin{equation}
S^{(1)}\symbol{126}\int L^{P}=\int L  \label{ad}
\end{equation}

String-local $L,V_{\mu }$ pairs are the starting point for the perturbative
construction of the $e$-independent S-matrix and correlation functions of
string-local fields. The problem how to maintain string-localization in
higher order perturbations is closely related to the problem of preserving
the $e$-independence of the S-matrix. This leads to a normalization
condition on higher order time ordered products of $L-\partial V~$which will
be commented on in section 4.

Note that the $L,V$ formalism is not directly applicable to $m=0\,$~since $%
\phi ~$and $V_{\mu ~}$have no massless limit. Behind this formal problem
there is a radical conceptual change (breakdown of Wigner-Fock particle
Hilbert space, infraparticles in QED, QCD confinement) whose spacetime
implications have remained outside of our conceptual understanding of QFT.
In fact these problems are outside the physical range of gauge theory;
whereas short distance properties of unphysical gauge dependent fields are
believed to share their asymptotic behavior with those of their sl physical
counterparts (in particular the QCD asymptotic freedom) one does not expect
that problems related to confinement can be accounted for in gauge theory.
Here the long distance fluctuations of string directions are expected to
become important. Following Wightman's reconstruction theorem \cite{St-Wi}
the massless QFT should be reconstructed from the massless limit of the
massive correlation functions thus avoiding direct questions concerning the
fate of the Wigner-Fock particle space.

\section{Wilson loops, Haag duality and the Aharovov-Bohm effect}

The following section extends ideas which were already presented in \cite%
{vector}.

Consider the spacelike Wilson loop for a string-local vector potential. In
the massive case one obtains from (\ref{lin})%
\begin{equation}
\doint A_{\mu }(x,e)dx^{\mu }=\doint (A_{\mu }(x)+\partial _{\mu }\phi
(x,e))dx^{\mu }=\doint A_{\mu }(x)dx^{\mu },~m>0
\end{equation}%
whereas in the massless limit the separate contributions to the integrand
diverge and instead one finds 
\begin{equation}
\doint (A_{\mu }(x,e)-A_{\mu }(x,e^{\prime }))dx^{\mu }=\doint \partial
_{\mu }(\phi (x,e)-\phi (x,e^{\prime }))dx^{\mu }=0~~for\text{ }m=0
\end{equation}%
It is important to notice that, although neither $\phi $ nor $\partial _{\mu
}\phi ~$possess massless limits, the mass singularities cancel in the
difference between $\phi ^{\prime }s$ with different $e$-directions. This
can be seen either in terms of the $e$-dependence of the intertwiners or by
using the fact that the $m^{-2}$ in (\ref{2-point}) cancel in the
2-pointfunction of the difference%
\begin{equation}
\left\langle \psi (x;e_{1},e_{1}^{\prime })\psi (x;e_{2},e_{2}^{\prime
})\right\rangle ,~~\psi (x;e,e^{\prime }):=\phi (x,e)-\phi (x,e^{\prime })
\end{equation}%
The $e$-independence of the loop integral despite its $e$-dependent
integrand is reminiscent of its gauge invariance in the Krein space setting
of point-local vector potentials. Later we will return to this analogy.

For the following it is convenient to work with operators instead of the
singular quantum fields. A regularization of the vector potential in terms
of a convolution with a smooth function $f~,$which is localized around a
small ball $B$ at the origin, leads to the regularized loop operator

\begin{equation}
\doint A_{\mu }^{reg}(x,e)dx^{\mu },~~A_{\mu }^{reg}(x,e):=\int
f(x-x^{\prime })A_{\mu }(x^{\prime },e)d^{4}x^{\prime }
\end{equation}%
It commutes with all operators whose localization region $\mathcal{O}$ is
such that there exists a direction $e$ for which the regularized
half-cylinder does not intersect $\mathcal{O}$. This includes in particular
all \textit{convex} regions which do not intersect the torus $l^{reg\text{ }%
} $which results from regularizing the loop $l.$

Operators whose localization region is such that there exists no choice of $%
e~$which permits to avoid an intersection with the regularized semi-infinite
cylinder$~l^{reg}+\mathbb{R}_{+}e$ do not commute with the regularized
Wilson loop. This includes in particular operators which are localized in a
torus which loops through $l^{reg~}$without intersecting it.

By allowing the $e~$to vary along the Wilson loop such that $e(\alpha )$
moves through a loop on the directional de Sitter spaces as $x(\alpha )~$%
sweeps through the Wilson loop, one enlarges the possibilities of avoiding
intersections; but$~$in case of a torus which intertwines $l^{reg}$ without
touching, an intersection with the $e$-extended Wilson loop is unavoidable.
The dependence on $e$ is "topological"; the Wilson "remembers" that its
integrand had a directional dependence but it forgets in which direction it
pointed.

This problem can be investigated directly in terms of the localization
property of the electromagnetic field strength $F_{\mu \nu }~$without using
vector potentials \cite{LTR}. The result is that the operator representing a
regularized magnetic flux through a surface $D~$does not change under
deformations of $D$ as long as its boundary $\partial D~$stays the same. Any
operator which is localized in a contractible region outside the regularized
torus $\partial D+B$ commutes with the flux operator but, as shown in \cite%
{LTR}, there are operators associated with interlocking but not intersecting
toroidal regions which do not commute with the regularized magnetic flux
operator. The authors refer to such a situation as the "breakdown of Haag
duality".

Recall that Einstein causality states that two operators commute if their
localization regions are spacelike separated. In terms of operator algebras
this means%
\begin{eqnarray}
\mathcal{A(O}) &\subseteq &\mathcal{A(O}^{\prime })^{\prime
},~~Einstein~causality \\
\mathcal{A(O}) &=&\mathcal{A(O}^{\prime })^{\prime },~~Haag~duality
\end{eqnarray}%
where the dash on the region refers to the causal complement and that on the
operator algebra to its commutant. Our intuitive understanding of causal
localization is however in terms of Haag duality \cite{Haag}; we expect that
an operator which commutes with \textit{all }algebras which are localized in
the causal complement of a region $\mathcal{O}$ is really localized in $%
\mathcal{O}~$ i.e. is a member of $\mathcal{A(O})$.

Haag duality holds for all algebras which are generated by \textit{massive}
free fields and is believed to remain valid for observable subalgebras
localized in multiply connected regions. But the properties of magnetic
fluxes in QED show that Haag duality is broken for multiply connected
subalgebras generated by point-local $s\geq 1~$massless field strengths ($%
F_{\mu \nu },R_{\mu \nu \lambda \kappa },..$). In other words there are
operators in $\mathcal{A(O}^{\prime })^{\prime }$ which do not belong to $%
\mathcal{A(O})$ and the Wilson loop operator with its topological $e$%
-dependence is an example.$~$Interestingly these breakdowns of Haag duality
happen in theories in which the potentials are necessarily string-local.

The "quirky" feeling that there may be some problems with causality in the
A-B effect \footnote{%
In most articles on the A-B effect the reader is assured that they are
unfounded, but they certainly play a role in its popular appeal.} has its
origin in the naive identification of the gauge theoretic quantum causality
in Krein space with that of a QFT in Hilbert space. The use of pl vector
potentials in Krein space is not incorrect; but it carries the danger of
identifying the Wilson loop as localized on a circle. The string.ls setting
is safe in this respect since the topological memory of the $e$-dependence
is precisely what one needs to be reminded of namely that, although the
Wilson loop object commutes with all operators localized in the causal
complement of the torus (the thickened Wilson loop), it is not localized on
it. This is what the breaking of Haag duality in Einstein-causal QFT means.
It occurs in all massless models involving massless $s\geq 1$ fields.

The violation of Haag duality is basically a classical phenomenon. It is
well-known that commutation properties of free quantum fields correspond to
"symplectic orthogonality" of their corresponding wave functions%
\begin{equation}
iIm(f,g)=\left[ A(f),A(g)\right] ,~f,g\text{ }real~test~functions
\end{equation}%
Hence the quantum A-B effect passes to its classical counterpart; the
correct classic vector potential is simply the expectation value of its
quantum counterpart in a suitable coherent state. The Stokes theorem does
not contain informations about \textit{physical} localization properties of
vector potentials.

The main reason for calling the readers attention to these facts (well-known
among experts) is that the unphysical aspects of the quantum gauge formalism
are not limited to problems of positivity (unitarity) but they also affect
the foundational causal localization principles. The correct localized
charge-carrying operators are obtained by smearing directional extended
Wightman fields $\Psi (x,e)$ with compactly supported test functions in $x$
and $e$.

The Hilbert space description of massless vector-potentials is traditionally
presented in the form of Coulomb (or radiation) gauge. Being the unique
Hilbert space potential which is rotation invariant in the $t=0$ hyperplane,
it is not surprising that it is obtained from integrating the string-local
potential over all string directions $e$ in the $t=0\ $hyperplane. As a The
lack of covariance and localilty prevents its application in renormalized
perturbation theory, but does not impede its use in quantum mechanics.

The phenomenon of breakdown of Haag duality is a general property of all
zero mass higher spin fields. For $s=2$ there are two string-local
candidates which can be viewed as the analogs of the string-local $A_{\mu }$
namely the string-local \thinspace $g_{\mu \nu }(x,e)$ (\ref{g}) or the
string-local $3$-tensor $A_{\mu \nu \kappa }(x,e)~$which results from a line
integration of the field strength $R_{\mu \nu \kappa \lambda }$ (\ref{R}).
The latter plays the analog role to that of the vector potential for $s=1$
in the verification of the breakdown of Haag duality. 

It is an interesting question whether in the Platonic world of Haag duality
violation there exists a relation between the multiple connectivity (the
genus) of the spacetime localization region and the spin of the string-local
zero mass potential. It also would be interesting to understand in what
sense the construction of covariant string-local potentials can be viewed as
a special case of recent constructions in \cite{Ruzzi}.

The important role of positivity for infrared aspects of the QED Hilbert
space can be seen by looking at simpler infrared problems in two-dimensional
models. The simplest such model is the derivative coupling $\bar{\psi}\gamma
_{\mu }\psi \partial ^{\mu }\varphi ~$of a $d=1+1~$fermion to the derivative
of a $m>0~$scalar massive field $\varphi ~$\cite{infra}. It has a solution
in terms of a formal exponential expression\footnote{%
For reasons of brevity we omit the Wick-ordering in field products at the
same point.}%
\begin{eqnarray}
&&\psi (x)=e^{ig\varphi }\psi _{0}(x),~\left\langle e^{ig\varphi
(x)}e^{-ig\varphi (y)}\right\rangle =\exp g^{2}i\Delta
^{+}(x-y),~\left\langle e^{ig\varphi (x)}e^{ig\varphi (y)}\right\rangle
=\exp -g^{2}i\Delta ^{+}(x-y)  \label{ex} \\
&&\exp g^{2}i\Delta ^{+}(\xi )=F(\xi ^{2}m^{2})\overset{m\rightarrow 0}{%
\rightarrow }(\xi ^{2}m^{2})^{-g^{2}},~~\phi :=\lim_{m\rightarrow
0}m^{g^{2}}e^{ig\varphi },~~\left\langle \phi \phi ^{\ast }\right\rangle
\neq 0,~\left\langle \phi \phi \right\rangle =0
\end{eqnarray}%
Although $\varphi ~$itself has no positivity-preserving massless limit, the
two-point correlation functions of its exponential field$~\phi ~$remain
finite (after rescaling with a $g$-dependent mass factor) and and are
consistent with conservation of the "$\phi $-charge". In fact the
combinatorial structure of the n-point function of $\phi $ in terms of
exponential $i\Delta ^{+}(x_{i}-x_{k})$ contraction reveals that $\phi $ is
a "$g$-charge" conserving field in the Hilbert space which the Wightman
reconstruction theorem associates with the limiting $\phi $ vacuum
expectation values \cite{St-Wi}.

There are two aspects of this construction which are worth mentioning.
Whereas in the massive case the Hilbert space of the full model was a tensor
product of Wigner-Fock $\varphi $-particle space with a $\psi _{0}$\
particle space this structure gets lost in the massless limit since the $g$%
-charge creating $\phi $\ field is not a member of the $\varphi $-Hilbert
space. rather $\phi $ creates charge sectors in which the charge-neutral $%
\partial \varphi $ (the limiting $\varphi $ does not exist) acts. Since the $%
\phi $ always appears together with the $\psi _{0},$ the Hilbert space is
actually a subspace of the tensor product of the $\psi _{0}$ with the $\phi $%
-space, i.e. the $g$-charge of $\phi $ coalesces wth the global $\psi _{0}~$%
charge. This in turn leads to a kind of kinematical infraparticle structure
which manifests itself in the absence of the mass-shell delta function; the
representation of the Poincar\'{e} group in the Hilbert space created by the
application of $\psi ^{\prime }s$ to the vacuum contains no discrete one
particle state but instead of a mass-shell delta functions one finds a
weaker threshold like singularity with a cut structure.

Suppose we ignore the $m\rightarrow 0~$limiting structure of exponentional
and define a free $\varphi _{0}$ function logarithmic two-point function%
\begin{equation*}
\left\langle \varphi _{0}(x)\varphi _{0}(y)\right\rangle =\log \mu
^{2}(x-y)^{2}
\end{equation*}%
In this case the formally associated free field violates positivity i.e. the
correlation functions define a linear indefinite metric space. In this case
there is no charge superselection structure. The situation resembles that of
the use of the Krein space point-local vector potential except that gauge
theory restores perturbative positivity for certain (gauge-invariant)
operators. But a structural understanding of problems behind infrared
divergencies of pl charge-carrying operators within GT is not possible. In
the SLFT setting \textit{the singular pl siblings }$\psi (x)$\textit{\
disappear} in the massless limit and the Maxwell charge-carrying fields
fields exist only as sl fields $\psi (x,e).$ At this point the "fake" matter
fields $\psi ^{K}(x)~$become fully misleading since the short distance
compensation from negative metric contribution feigns a non existing pl
localization; in this case the Hilbert space positivity does not even allow
the existence of the singular pl siblings of the $\psi (x,e).$

The understanding of long range properties of electric charges and the
infraparticle aspect in QED pose more demanding dynamic challenges which go
far beyond the kinematical observations on two-dimensional models. There
remains however a formal analogy with properties one expects in the sl
Hilbert space formulations. In order to highlight these analogies it is
helpful to reformulate the previous observation by viewing the massless $%
\phi $-fields as limits of spacelike line-integrals. With $j_{\mu }=\partial
_{\mu }\varphi $ we may define the charge-carrying field $\phi $ directly in
the massless model

\begin{equation}
e^{-ig\varphi }=lim_{\Lambda \rightarrow \infty }e^{ig\int_{0}^{\Lambda
}d\lambda j_{\mu }(x+\lambda e)e^{\mu }},~e^{2}=-1  \label{an}
\end{equation}%
This suggests the following analogy ($\tilde{j}^{\mu }=\varepsilon ^{\mu \nu
}j_{\nu }$)%
\begin{eqnarray}
&&\exp ig\varphi (x,e)\sim \exp ig\Phi (x,e,e^{\prime }):=\Psi  \label{ana}
\\
with~\varphi (x,e) &=&\int_{0}^{\infty }j_{\mu }(x+\lambda e)e^{\mu
}d\lambda \text{ }and\text{ }\Phi (x,e,e^{\prime })=\int_{0}^{\infty }A_{\mu
}(x+\lambda e^{\prime },e)e^{\prime \mu }d\lambda  \notag
\end{eqnarray}%
In both cases the fields are logarithmically infrared divergent in the
massless limit, whereas the exponential operators (defined as above by
Wightman reconstruction from the massless limit of vacuum expectation
values) remain finite.

In analogy to the $\phi $ charge we would like to view a state created by$%
~\Psi $ as carrying a Gauss charge i.e.

\begin{equation}
Q\left\vert \Psi \right\rangle :=\int d^{3}x\vec{\nabla}\vec{E}\left\vert
\Psi \right\rangle =\lim_{S\rightarrow \infty }\doint\limits_{S}\vec{E}d\vec{%
S}~\left\vert \Psi \right\rangle =\lim_{S\rightarrow \infty
}\doint\limits_{S}\left[ \vec{E},\Psi \right] d\vec{S}\left\vert
0\right\rangle \neq 0~?
\end{equation}%
Clearly such a superselected state cannot be compactly localized. Using
exponential line integrals over point-local gauge potentials fails, since
indefinite metric is not compatible with charges superselection rules,
whereas the above ansatz has a better chance. In analogy to the 2-dim. model
the full one-electron state should be of the form $\psi _{0}(x)\Psi
(x.\infty )\left\vert 0\right\rangle $ where $\psi _{0}~$is the free
electron field and $x$ in $\Psi ~$refers to the start of the semi-infinite
strings.

One of the few rigorous results in QED is a theorem that the Lorentz
symmetry is spontaneously broken in sectors of nontrivial electric charge 
\cite{Froe}. This certainly does not happen in interactions with massive
vector mesons. The heuristic picture is that the strings of charged
particles are the centers of regions of noncompact infrared photon clouds.
This is consistent with the established fact that such photon clouds leads
to continuously many directional superselection sectors within a fixed
charged sector \cite{Haag}. Whether the above $\Psi $-states have this
property can be checked by studying the energy-momentum density between such
states.

Returning to the question of the structure of the Hilbert space one may
summarize the present situation as follows. Independent of the pl or sl
field localization the Hilbert space of asymptotically complete theories
with a mass gap is a Wigner-Fock particle space where the particles are
related to the interacting fields by LSZ scattering theory. This is a very
clear conceptual situation. In massless limits involving sl $s\geq 1~$%
potentials this picture breaks down; in such cases the structure of the
Hilbert space is expected to involve massless limits of nonpolynomial (in
the case of QED exponential ) string-local new sector creating composites of
massless potentials. In problems with a mass gap one expects that the
Wigner-Fock space provided by scattering theory to be complete ("asymptotic
completeness").

The reconstruction of a Hilbert space from massless limits of correlation
functions in terms of suitable limits of free fields is expected to require
the use of exponential non-local limits of free fields as in (\ref{ana}$\ $%
which create inequivalent representation spaces. But even in case the above
proposal to describe the Hilbert space of QED turns out to be correct, there
remains the problem how these inequivalent representations generating
descriptions are spacetime-related to the interacting fields. Such
constructions are expected to lead to a better understanding of the momentum
space recipes \cite{BN} \cite{YFS} in terms of spacetime concepts of
collision theory may be helpful. 

Morchio and Strocchi addressed the infrared problem within the gauge
theoretic setting by constructing positivity obeying topology on the Krein
space formalism \cite{M-S}. We believe that the understanding of such
long-distance problem will be simpler and more natural in a description
where these string-local objects are already part of the perturbative
setting and the topology of the Hilbert space is already the correct one. 

A intriguing proposal can be found in a recent article by Buchholz and
Roberts \cite{Bu-Ro}. These authors observe that a restriction of the
Minkowski space to a forward light cone $V_{+}$ would still permit a
complete description of QFTs with a mass-gap, but its irreducibility would
be lost in the presence of photons. In the context of string-local fields
this situation suggests to use fields localized on spacelike hyperbolic
curves which stay inside $V_{+}$ and only touch the surface of the light
cone at lightlike infinity. Such a situation could lead to a more natural
way to implement infrared cut-offs. What is missing is a perturbative
realization of this idea; but the increasing perturbative experience with
fields localized on spacelike or lightlike\footnote{%
I am indebted to Jens Mund for informing me that the use of lightlike
instead of spacelike linear strings causes no new problems.} lines suggests
that such an extension may be possible.

\section{Differential-geometric control of directional fluctuations}

Whereas setting up first order string-local interactions in the form of $%
L,V_{\mu }$ pairs within the power-counting restriction is basically a
kinematic problem involving free fields, the situation changes when it comes
to the construction of the higher order S-matrix. The reason is that the
singular nature of time ordering does not permit to take the divergence $%
\partial ^{\mu }T(..\partial _{\mu }\varphi ..)$ directly through the
time-ordering to the affected operator. The differential relation\footnote{%
Unpublished remark \ by Jens Mund (Vienna 2011).} 
\begin{eqnarray}
&&d(TLL^{\prime }-\partial ^{\mu }TV_{\mu }L^{\prime }-\partial ^{\prime \mu
}TLV_{\mu }^{\prime }+\partial ^{\mu }\partial ^{\prime \mu }TV_{\mu }V_{\mu
^{\prime }}^{\prime })=0,~~d=d_{e}+d_{e^{\prime }}  \label{indep} \\
&&TLL^{\prime }|^{P}:=TLL^{\prime }-\partial ^{\mu }TV_{\mu }L^{\prime
}-\partial ^{\prime \mu }TLV_{\mu }^{\prime }+\partial ^{\mu }\partial
^{\prime \mu }TV_{\mu }V_{\mu ^{\prime }}^{\prime },  \notag
\end{eqnarray}%
which secures the $e$-independence of the second order S-matrix in the
adiabatic limit (formally the integral over Minkowski spacetime)\footnote{%
In massive theories boundary terms at infinity vanish. Without x,x'
integration the expression in the bracket can be used as a definition of a
second order point-local interaction density..}, would be a trivial
consequence of (\ref{p}) if it were not for singularities in $T$-products
from coalescent points and crossing of strings.

In terms of differential forms in the de Sitter space of directions the
individual contributions to the right hand side are zero forms and their sum
is an exact form. We remind the reader that in the presence of a mass gap
there are no boundary terms at infinity, so that in analogy to (\ref{ad})
the divergence terms do not contribute to the second order S-matrix%
\begin{equation}
S^{(2)}\symbol{126}\int \int TLL^{\prime }|^{P}=\int \int TLL^{\prime }
\label{S}
\end{equation}

It turns out that such "normalization" problems as posed by (\ref{indep})
can be solved by using the freedom in defining time-ordered products.
Starting from a "kinematic" time ordering $T_{0},$ one computes the anomaly $%
A$ as the singular part of the terms containing derivatives

\begin{equation}
-A=s.p.(-\partial ^{\mu }T_{0}V_{\mu }L^{\prime }-\partial ^{\prime \mu
}T_{0}LV_{\mu }^{\prime }+\partial ^{\mu }\partial ^{\prime \nu }T_{0}V_{\mu
}V_{\nu }^{\prime })  \label{3}
\end{equation}%
Since in the following we will be interested in the S-matrix, we only need
the contribution from the 1-contraction (the tree component) $A|_{1-contr.}$%
; for notational economy we will omit the subscript, so in the following
relations the $A$ stands for the tree component of the anomaly.

The kinematic $T_{0}$ is defined by taking all derivatives outside e.g. 
\begin{eqnarray}
\left\langle T_{0}\partial \varphi \partial ^{\prime }\varphi ^{\ast \prime
}\right\rangle &:&=\partial \partial ^{\prime }\left\langle T_{0}\varphi
\varphi ^{\ast \prime }\right\rangle \\
\partial ^{\mu }\left\langle T_{0}\partial _{\mu }\varphi \varphi ^{\ast
\prime }\right\rangle &=&-i\delta (x-x^{\prime })-m^{2}\left\langle
T_{0}\varphi \varphi ^{\ast \prime }\right\rangle  \notag
\end{eqnarray}%
If the anomaly is of the form 
\begin{equation}
A=-N+\partial ^{\mu }N_{\mu }+\partial ^{\mu }\partial ^{\nu }N_{\mu \nu }
\end{equation}%
where the $N^{\prime }s$ contain $\delta (x-x^{\prime })$ functions, they
can be absorbed as renormalization terms in the time-ordered operator
products (\ref{indep}). They describes the delta function terms which
violate the relation (\ref{indep}) if one uses the kinematical $T_{0};~$%
hence the anomaly terms reveal in what way the time ordered products at
coalescent points (or more general at string intersections) in (\ref{indep})
have to be defined.

For the calculation of the second order S-matrix one only needs to compute
the $N.~$Note that $N~$terms are similar to the counterterms well known from
the renormalization formalism for point-local interactions. But there is a
significant conceptual difference. Whereas the renormalization counterterms
in point-local renormalization come with new coupling parameters, the
contact terms originating from anomalies are uniquely determined in terms of
the basic first order couplings and the masses of the free fields in terms
of which the first order interaction density is defined. Such anomaly terms
will be referred to as \textit{induced} \textit{interactions}. They
originate \textit{from the implementation of the }$e$\textit{-independence
of the S-matrix} and hence they have no analog in $s<1$ point-local
interactions.

The calculation of the S-matrix requires only the calculation of $N.$ In
that case it is more convenient to use a weaker formulation%
\begin{eqnarray}
&&dL-\partial ^{\mu }Q_{\mu }=0,~Q_{\mu }:=d_{e}V_{\mu }~  \label{Q} \\
&&dTLL^{\prime }-\partial ^{\mu }TQ_{\mu }L^{\prime }-\partial ^{\prime \mu
}TLQ_{\mu }^{\prime }=0  \notag
\end{eqnarray}%
As will be seen later, this "$Q$-formulation" is closely related to the
implementation of the gauge invariant S-matrix in the $sS=0$ in the CGI BRST
setting.

The $V$-formalism, which leads to the definition of pointlocal interactions
densities (\ref{indep}), turns out to be indispensable for the construction
of interacting string-local fields. It permits to define higher order
point-local interaction densities in terms of the renormalizable
string-local formalism. Together with the formal relation between point- and
string-local matter fields (\ref{psi}) it can be used for the perturbative
construction of renormalized correlation functions of string-local fields.
The point-local higher order interaction densities (\ref{indep}) play the
role of the $e$-independence of string-local field correlations from the $%
e^{\prime }s$ of internal propagators. This construction of interacting
string-local fields has been initiated by Jens Mund and will be the subject
of a forthcoming publication.

Assuming that these string-local fields remain renormalizable (Wightman
fields in $x$) in every order, the results of Jaffe \cite{Jaffe} on the
singular nature of exponentials of scalar free fields suggest that $\psi
^{P}(x)=\psi (x,e)e^{-ig\phi }$ is a singular (not Wightman-localizable)
field in a well-defined theory. As previously mentioned it shares its bad
short distance properties (unbounded increase of $d_{sd}$ with the
perturbative order) with that of the field in the point-local setting, but
at least its singular behavior is not accompanied by a (with perturbative
order) growing number of coupling parameters. The bad high enery behavior of
the off-shell vacuum expectation of the nonrenormalizable pl Hilbert space
setting remains, but the numerical coefficients of the worsening pl
counterterms do not introduce new parameters.

In this situation one expects that there are high energy on-shell
cancellations which permit the scattering amplitudes of the singular pl
fiels to be the same as those obtained from the renormalizable sl
correlations. In such models the cause of pl nonrenormalizability is a
weakening of localization. Assuming that these observations are correct, the
lack of renormalizability either indicates a weakening of localization or
the PCB violating coupling does not define a QFT.

Another important new concept which has no analog in $s<1~$pl interactions
is the before-mentioned appearance of \textit{induced terms. }In the
following this will be illustrated in terms of sl second order calculations
three different models.

The first such model is scalar massive QED (\ref{scalar}). In the case the
anomaly contribution arises from the divergence acting on the two point
function involving a $\partial _{\mu }\varphi .~$The result for the induced
contact contribution is the expected second order quadratic in $A_{\mu }$
contribution 
\begin{equation}
g^{2}:\varphi ^{\ast }(x)A_{\mu }(x,e)\delta (x-x^{\prime })\varphi
(x^{\prime })A^{\mu }(x,e^{\prime }):+~(e\leftrightarrow e^{\prime })
\label{Wick}
\end{equation}%
which may be absorbed into a change of $T_{0}\rightarrow \NEG{T}$ product of
the $\partial \varphi \partial ^{\prime }\varphi ^{\ast \prime }~$%
contraction contributing 
\begin{equation*}
\left\langle T_{0}\partial _{\mu }\varphi ^{\ast }\partial _{\nu }\varphi
\right\rangle \rightarrow \left\langle T\partial _{\mu }\varphi ^{\ast
}\partial _{\nu }\varphi \right\rangle =\left\langle T_{0}\partial _{\mu
}\varphi ^{\ast }\partial _{\nu }\varphi \right\rangle +cg_{\mu \nu }\delta
(x-x^{\prime })
\end{equation*}
For more details we refer to \cite{vector}.

This is similar to gauge theory where it results from BRST gauge invariance
of the S-matrix\footnote{%
In the naive formulation the quadratic term arises from the replacement of $%
\partial \rightarrow \partial -igA$ in order to preserve gauge invariance of
the classical Lagrangian.} $\mathfrak{s}S=0,~$except that the independence
of the S-matrix from string directions is a natural physical consequence of
large time scattering theory \cite{Bu-Fr} for string-local fields in the
presence of a mass-gap. In a perturbative setting, which is based on the
construction of the S matrix in terms of the adiabatic limit of Bogoliubov's
formal time-ordered products of interaction densities, it has to be imposed (%
\ref{indep}).

There is a fine point which turns out to be of significant conceptual
importance. The directional fluctuations only disappear after adding up all
contributions to a particular scattering amplitude to the same perturbative
order. For the case at hand the singularity at $e=e^{\prime }~$in the
time-ordered propagator (which enters the second order tree contribution to
scattering) corresponding to the two-point function (\ref{2-point}) 
\begin{equation*}
\frac{1}{p^{2}-m^{2}-i\varepsilon }(-g_{\mu \mu ^{\prime }}-\frac{p_{\mu
}p_{\mu ^{\prime }}(e\cdot e^{\prime })}{(p\cdot e-i\varepsilon )(p\cdot
e^{\prime }+i\varepsilon )}+\frac{p_{\mu }e_{\mu ^{\prime }}}{(p\cdot
e-i\varepsilon )}+\frac{p_{\mu }e_{\mu ^{\prime }}^{\prime }}{(p\cdot
e^{\prime }+i\varepsilon )})
\end{equation*}%
is ill-defined since the two fluctuating directions enter with a different $%
\varepsilon $-prescription\footnote{%
This is the reason why the axial gauge interpretation failed.}. But in the
use of this propagator for the second order \textit{on-shell} scattering
amplitude ( the scalar analog of M\o ller- and Bhaba-scattering) this
problem disappears. This is similar to the verification of on-shell gauge
invariance in the second order tree approximation except that $e,e^{\prime
}~ $are~not global gauge parameters in an Krein space gauge theory but
rather spacelike directions of independently fluctuating string-local vector
potentials acting in a Hilbert space.

Whereas in the Krein space gauge setting individual perturbative
contribution to a scattering process exists regardless whether it is gauge
invariant by itself or not, the independently fluctuating $e^{\prime }s~$%
become only "mute" after adding up sufficiently many perturbative on-shell
contributions in a fixed order. Any attempt to interpret the independently
fluctuating $e^{\prime }s$ as gauge parameters by equating them in
contributions from different fields causes the renormalization-resistent
infinities of the abandoned axial gauge formalism to return with full
vengeance.

It is important to notice that the second order $A\cdot A\varphi ^{\ast
}\varphi ~$contribution is induced by the principles of QFT (causal
localization in Hilbert space), there is no reference to a classical gauge
structure. This is even more remarkable in models of self-interacting vector
mesons in which the Lie-algebra structure, unlike classical gauge theory,
bears no relation to an imposed symmetry but rather results from only
implementing the spacetime causality principles. That the classical
causality principles of Faraday, Maxwell and Einstein become such powerful
in the context of Hilbert space positivity of interacting higher spin $s\geq
1$ quantum matter is truly surprising.

A much richer second order induction occurs in the $gA\cdot AH~$coupling of
massive vector mesons to a Hermitian matter field $H~$(\ref{H})$.$ In this
case the requirement of second order $e$-independence (\ref{indep}) of $S$
for the more involved first order $L,\partial V$ leads to a larger number
(no charge symmetry) of induced interactions which includes the $H^{4}$
self-interaction and even requires the presence of a first order
modification by $H^{3}~$self-interactions. The sum of both contributions has
the form of a Mexican hat \cite{vector} just as in the CGI setting in \cite%
{Scharf}, where the numerical coefficients contain besides the coupling
strength $g~$also mass ratios of the two physical masses $m_{H},m.$ An
unsettled point at the time of wrinting will hopefully be addressed in a
separate paper.

These\textit{\ induced} "Mexican hat" terms would also appear in the BRST
gauge formalism but, as already mentioned before, in a diagrammatic Feynman
formalism it is difficult to be aware of the differences between on- and
off-shell properties. Implementing the on-shell BRST $\mathfrak{s}$%
-invariance $\mathfrak{s}S=0,$ or the $e$-independence $dS=0~$in the SLF
Hilbert space setting, it is impossible to overlook the \textit{induction}
of a Mexican hat like self-interaction or confuse it with an off-shell SSB
mechanism; it is the massive vector meson which is in the driver's seat and
not the scalar matter coupled to it \cite{Scharf}. Apart from the richer
structure of the induced Mexican hat potentials (related to the absence of
the charge selection rule for $H$-couplings), this is analog to the second
order induction of the quadratic $A$-dependence\footnote{%
There is no reason to refer to the $\partial \rightarrow \partial -ieA$
classical \ gauge connection, the quadratic $A~$term results solely from the
quantum principles of QFT..} in massive scalar QED. In both cases the only
conserved current is the "massive Maxwell current" $j_{\mu }^{Max}=\partial
^{\nu }F_{\mu \nu }$ and its charge is always screened ($Q^{Max}=0$) and not
spontaneously broken ($Q^{SSB}=\infty $)~\cite{vector} (and references
therein).

It is interesting to note that unitarity properties of tree graphs and a bit
of common sense, ignoring classical gauge structures, led some authors \cite%
{Tik} \cite{Corn} to the correct result already in the early 70's; including
the idea that the Lie-algebraic structure of self-interacting vector-mesons
may be a natural property of self-interacting massive vectormesons and needs
no imposition of gauge symmetry. If they had formulated their unitarity
requirements directly in terms of interactions of massive vector mesons with
Hermitian matter fields they would have noticed that there is no need to
invoke a SSB Higgs mechanism. Pushing a bit harder they could have seen that
the reason why massive QCD or just Y-M (as opposed to massive QED) requires
the presence of a $H$ coupling just in order to avoid violations of the
second order PCB (and not for implementing SSB). The present work on
interactions of $s=1$ vector mesons in a Hilbert space setting can be seen
as a field theoretic confirmation of these on-shell unitarity arguments
which need neither the invocation of gauge symmetries nor that of SSB.

It is certainly true that the formalistic "breaking of gauge invariance" by
a shift in field space of scalar QED combined with an hermiticity-restoring
gauge transformation\ leads to the abelian Higgs model (what else could it
do?). \textit{But this bears no relation to the intrinsic properties of the
resulting model}.

In any renormalizable model the form of the local interaction density is
uniquely determined in terms of the field content (including the physical
masses of the free fields) and the imposed internal symmetries.
Renormalization theory leads to additional contributions in the form of
counter-terms with new coupling parameters. The new insight is that the
preservation of Hilbert space positivity in the presence of $s=1\ $%
vector-potentials leads to the phenomenon of \textit{induced} interactions;
in contrast to the free parameters of counter-terms (which are only subject
to the imposed inner symmetries), \textit{induced contributions do not
enlarge the number of free parameters}. In particular the numerical
coefficients of the induced Mexican hat potential are determined in terms of
the $A\cdot AH$ coupling strength $g$ and the two masses $m,m_{H}.$

SSB in the sense of Goldstone ($\partial ^{\mu }j_{\mu }=0\,\ $but $Q=\infty 
$) requires an \textit{interaction density which contains massive as well as
massless scalar free fields in a precisely tuned proportion}. A helpful way
to obtain such a situation is to start from the conserved current of a
symmetrically coupled self-interacting scalar multiplet with a symmetric
mass term of the opposite sign (so the potential takes the form of a Mexican
hat). It is well known that a suitable shift in field space brings the
minimum back to the value zero. One then checks that this shift
manipulations changes the form of the field equation but the conservation of
the current (which uses the field equation) remains intact. The new field
equations contain mass terms which depend on the choice of the direction of
the field shift in the multi-component field space, but there always will be
some zero mass particles (the Goldstone bosons).

After the semiclassical preparation of the model, QFT (in the form of
renormalized perturbation theory) takes over. It maintains the physical
masses of the interaction density defining free fields and the conservation
of the current in every order of perturbation theory. Some of the integrals
defining the conserved charges (the would be global symmetry generators) are
prevented from converging by long range manifestation of the Goldstone
bosons. This (and not the quasiclassical manipulations to obtain such a QFT)
is the \textit{definition of SSB} (the non-invertibility of the Noether
theorem i.e. $\partial j=0$ but $Q=\int j_{0}=\infty $). An alternative more
general picture is to say that there are "partial" charges localized in a
causally closed spacetime regions whose exponentiated unitary symmetry
operators implement the full symmetries on observables in a smaller
spacetime region. Some of these partial charges fail to be extendible to
charges which implement global unitary symmetries. For a model-independent
structural theorem about SSB see \cite{E-S}.

\textit{QFT is not a theory which says anything about the masses of
elementary free fields} which enter the definition of a model. Masses of
bound states interpolated by composite fields on the other hand are expected
to be computable (although there is no known consistent perturbative method
to implement this. The terminology "SSB creation of masses" is still within
the range of tolerable metaphors of particle physics. The danger of
misunderstandings starts if these manipulations are applied outside their
range of validity.

As explained previously for interactions involving higher spin fields there
are restrictions (gauge invariance, directional independence of $S$) which
lead to the induction of $H~$selfinteraction. Starting from $A\cdot AH~$%
interactions between a massive vector meson and a Hermitian field one
obtains the same Higgs models as from the SSB Higgs mechanism which starts
from a massless gauge theory of a complex field. Which is the correct
derivation i.e. the derivation which is consistent with the physical
principles? The answer is that the one based on the gauge invariance of the
S-matrix or even better on $e$-independence of $S$ is consistent with QFT
principles. The reason is obvious since SSB, even there where it is
meaningful is an option whereas gauge invariance or the properties of SLFT
are part of the conceptual structure of interacting massive vector mesons.

QFT is a foundational quantum theory which reveals its intrinsic properties
independent of by what computational tricks or thoughts the acting physicist
obtained a model. In case of SSB the integral over the charge density of the
conserved current diverges (the definition of SSB) and in models involving
massive vector mesons the charge is "screened" i.e. the $Q$ vanishes. This
is obviously a generic property of an identically conserved Maxwell current $%
j_{\mu }=\partial ^{\nu }F_{\mu \nu }$ \cite{vector}. One should not allow
to confuse such opposite situations by metaphors as: "photons fattened by
eating Goldstones".

One obtains a better understanding of the history of the Higgs model when
one recalls that QED resulted from a natural adaptation of the classical
Maxwell's theory to the requirements of charge-carrying (complex) quantum
matter. That classical theory contains however no suggestion of how to
couple vector potentials to \textit{Hermitian} matter. As we know nowadays
such couplings vanish in the limit $m\rightarrow 0$ i.e. they only exist in
the presence of massive vector mesons. Therefore the SSB prescription of a
shift in field space was a helpful formal device to become aware of a new
coupling to Hermitian matter which disappears in the Maxwell limit. The
problem started when this prescription was misinterpreted as a physical SSB
of gauge symmetry which "creates" masses. This does not only reveal a
misunderstanding of the role of the quantum gauge formalism (a method to
recover physical properties from an unphysical Krein space setting), but
also of SSB (a conserved current with a \textit{diverging} charge).

However in \textit{massive} \textit{QCD} or just massive \textit{Y-M} one 
\textit{really needs a coupling to a }$H$\textit{-field,} so the important
remaining problem (the most important of the entire section) is to
understand why for self-interacting massive vectormesons the consistency of
the model requires the presence of $H^{\prime }s,~$whereas in massive
(spinor or scalar) QED they play no role.

GT and the Hilbert space based SLFT lead to the same answer to this
question, though the details of the second order calculations are somewhat
different. The gauge theoretic requirement $\mathfrak{s}S=0$ \cite{Scharf},$%
~ $as well as the $dS=0$ formalism derived from first principles of QFT
induce nonrenormalizable second order self-interactions of $d_{sd}=5.$ 
\textit{This is the only known case in which a first order interaction
within the power-counting restriction }$d_{sd}^{int}\leq 4\,\ $\textit{turns
nonrenormalizable in second order! }When in the old days the 4-Fermion model
of weak interactions was replaced by the short-distance improved
intermediate massive vector meson model in the setting of massive nonabelian
gauge theory, it was at first overlooked that this model fails to preserve
second order renormalizability. Hence the massive vector meson exchange
converted Fermi's 4-Fermi coupling first-order PCB whereas the second order
compensation with $A$-$H$ interactions maintains this in second order.

The compensatory mechanism is reminiscent of compensations of short distance
singularities between contributions from different spins in supermultiplets.
The compensating field coupled to the self-interacting massive vector mesons
should have a lower spin (a higher spin would make sense worse) and the same
Hermiticity properties as the $A_{\mu }$, hence it must be a $H$-field.

It turns out that the only way to preserve the second order power-counting
restriction is to \textit{enlarge the first order }$A$\textit{%
~self-interaction by a nonabelian }$A\cdot AH$\textit{\ coupling\footnote{%
The short distance compensation does nit fix the number of $H^{\prime }s.$
In case of one $H$ (which seems to be favored by LHC) the coupling is unique.%
}}. Its second order contribution contains in addition to those term
expected from the analogy with the abelian model an additional $d_{sd}=5$
contribution which can be used to compensate the incriminated
nonrenormalizable term from the self-interaction. The net result is that,
apart from the induced Mexican hat potential, all non-compensating anomaly
contributions can be absorbed into modified time-ordered products
(propagators) $T_{0}\rightarrow T\,.~$There is no physical role for the $H$
field to play in the understanding of massive abelian gauge theories, but
their compensatory role in the presence of \textit{self-interacting} massive
vectormesons is indispensable.

This preservation of renormalizability by short distance compensations is
reminiscent of cancellations between different spins contributions in
supersymmetric interactions. The difficulty that one does not know how to
formulate a supersymmetry-preserving renormalization theory is absent in
case of gauge invariance or $e$-independence. The formal manipulations
within a "Higgs mechanism" give the complete interaction in one swoop but
fail to realize that behind the scene there is a new type of compensation
between different spin contributions which different from those of
supersymmetry. For the GT presentation of this compensation the reader
should consult \cite{Scharf}, some remarks about its SLFT derivation can be
found in \cite{vector} and the details will be contained in joint work with
Jens Mund.

The reason why it could be important to understand the correct physical
cause for the presence of the $H$ (apart from the fact that in a
foundational theory as QFT it is always good to know the intrinsic reasons
for a new phenomenon) is that the construction of $L,V_{\mu }$ pairs may
lead to compensatory situations involving higher spin fields as $g_{\mu \nu
}(x,e)~$together with compensatory $s=1$ and $s=0$ contributions. It would
not be necessary that $V_{\mu }$ has $d_{sd}\leq 4~$as long as the higher
short distance dimensions compensate in higher orders.

In the Higgs paper \cite{Higgs} one finds the remark that the $H$-field is
needed to generate a mass. In this context there are references to works in
condensed matter physics as the phenomenological Landau-Ginsberg description
of superconductivity and its microscopic BCS refinement (also Anderson's
work on energy gaps from a gap-less situation). In all those works already 
\textit{existing degrees of freedom regroup themselves} (the Cooper pairs)
and in this way lead to new physical manifestations. But this is precisely
the mechanism by which the Hilbert space description of \textit{massive} QED
requires the introduction of a scalar string-local field $\phi $ (the $%
A_{\mu }^{\prime }s$ "escort") which depends on the same degrees of freedom
(namely the $s=1$ Wigner creation/annihilation operators) as the massive
vector potential.

As the BCS superconductivity cannot be described without the regroupment of
condensed matter degrees of freedom in the form of Cooper pairs, massive QED
in a Hilbert space setting needs the presence of the (degrees of freedom
maintaining) escort field $\phi .~$Gauge theory in Krein space is a physical
incomplete theory (no physical substitute for gauge-dependent fields) which
has many additional unphysical degrees of freedom (negative metric fields,
ghost fields) but no physical $\phi $-field. The degree of freedom issue and
the appearance of additional fields is a quantum phenomenon which has no
classical counterpart. Whereas the perturbative structure of $s<1$ QFT
maintains formal relations to classical FT, these links are lost in the
presence of $s\geq 1$ fields.

Precisely for this reason it is important to understand the physical cause 
\textit{why additional }$H$\textit{-degrees of freedom are needed in the
presence of self-interacting massive vector mesons}. It turns out that the
first order modification of the 4-Fermi interaction by intermediate massive
vector mesons leading a nonabelian massive gauge theory only secured the
first order PCB but that the second order leads to a
renormalizability-violating $d_{sd}=5~$term. Since in all previous studied
models of QFT the first order PCB bound $d_{sd}\leq 4$ always implied
renormalizability\footnote{%
Higher order short distance compensation between fields of different spin
are only expected in SUSY QFTs.}, this phenomenon was overlooked. The
interpretation of the nonabelian Higgs model as resulting from a SSB
breaking of gauge invariance obscured this problem since the \textit{physical%
} understanding of a second order short distance compensation mechanism
cannot be delegated to formal description of a model as arising from SSB of
gauge symmetry (a physical symmetry ?). In the more physical SLFT Hilbert
space setting for $s=1~$interactions there is no gauge symmetry, so what is
the S to which the SSB prescription should be applied ?

One rather has to realize that the coupling of $H$ fields to massive vector
mesons defines autonomous theories and that the $H$-selfineractions (the
Mexican hat potential) is "induced" and not the result of an imposed SSB.
The induction phenomenon, which was explained in the beginning of this
section, does not exist for point-local $s<1$ interactions but is
characteristic for interactions involving string-local $s\geq 1$ fields.
Another phenomenon which appears only for $s\geq 1$ is that a massive model
within first order PCB may violate second order PCB. In such a case one has
to look for an extension of the model by adding an interaction with an
additional field which leads to a second order compensation of the
renormalizability threatening second order terms. Precisely this happens for
self-interacting massive vector mesons in which the compensating fields are $%
H$ fields; this, and not the alleged role of a mass creating midwife, is the
foundational purpose of the $H.$

As far as the classification of all second order compensation of
self-interacting massive vector mesons of different masses couples with $H$%
-multiplets with different mass is concerned the recipe based on SSB
breaking schemes turns out to be quite efficient because the SSB formalism
leads to the same family of models as the classification of compensations 
\cite{BDSV}. Nature does not follow the logic of ostensible simplicity of
calculational prescriptions of particle theorists but rather sticks to her
own principles. The relation of pertubative renormalizability with causal
localization in a Hilbert space setting is certainly one of her deep but
still insufficiently understood conceptual messages.

That important discoveries have been made by less than correct reasons is
not new; even Dirac's argument in favor of anti-particles was based on the
inconsistent hole theory. For the experimental discovery of the Higgs
particle this made no difference; one may even speculate that the somewhat
mysterious appeal of the SSB explanation (creating the masses of all
particles, including its own) may have been a more important motivating
force for the enormous mental and material effort which went into its
discovery at LHC than the preservation of second order renormalizability
(which to many particle physicists may appear as a d\`{e}j\'{a} vu of the
4-Fermi past).

What is a bit worrisome (as compared to the history of antiparticles and
other important discoveries originally made for less than correct reasons)
is that the SSB Higgs mechanism still dominates the scene after 4 decades.
This raises the question whether explanations which found acceptance within
a globalized community (apart from the tiny group of scholars who noticed
conceptual inconsistencies) can be still be corrected in times of "Big
Science". In any case the main purpose of the present work is not to address
sociological problems but rather the presentation of a surprising new look
at one of QFT's oldest problems: how to formulate $s\geq 1$ interactions in
a quantum theoretical Hilbert space setting.

This includes in particular problems involving massless vector potentials.
The perturbative infrared divergencies for scattering of electric
charge-carrying particles signal the breakdown of the spacetime
understanding of the field-particle relation in terms of (LSZ, Haag-Ruelle)
scattering theory. Successful momentum space recipes (the Bloch-Nordsieck
prescription and its Yennie-Frautschi-Suura refinement \cite{YFS}) for
photon-inclusive cross sections cannot hide the fact that GT, as a result of
absence of charge-carrying matter fields, is not able to solve this problem.

One expects that the new SLFT theory sheds new light on this age-old
problem. After all passing form point-local fields in GT to covariant
string-local interacting fields is not the result of a playful attitude to
try out something else (as in ST), but rather the inevitable consequence of
upholding one of quantum theory's most cherished principles: the positivity
of the Hilbert space.

The correct analog of quantum mechanical long-range Coulomb potentials are
string-local charge-carrying matter fields. The associated particles are
still registered in compact localized counters; the only influence
noncompact field localization has on particle counting events is that are
spread over a larger spacetime region. As previously mentioned there exist
rigorous results based on the quantum Gauss law which show the
noncompactness of charge localization and the "stiffness" of string-local
charge-carrying states whose directions (different from that of their
massive counterparts) cause a spontaneous breaking of Lorentz-invariance 
\cite{Bu} \cite{Froe} and leads to a radical structural change of the Fock
space (see end of previous section).

The only objects which remain relatively unaffected by the structural
changes in the massless limit are the expectation values of string-local
fields. Their calculation in the massive case requires the use of the
point-local interaction densities $(TL..L)^{P}$ (\ref{indep}) as well as the
relation (\ref{psi}) between point- and string-local matter fields. The
extension of the on-shell S-matrix formalism to off-shell string-local
fields is an ambitious program which has been initiated by Jens Mund. One
expects that its leads to both renormalizable (Wightman-localizable)
string-local and singular (Jaffe class) point-local matter fields. Only the
string-local correlations are expected to possess a massless limit; they
account for the properties of the "rigid" (SSB L-invariance-breaking)
electric charge-carrying strings.

The QED Hilbert space and the string-local operators acting in it are then
determined in terms of Wightman's reconstruction theorem \cite{St-Wi}. They
should provide the spacetime explanation for the change of the electron
mass-shell into a milder cut-like singularity and the momentum space recipes
which replace the vanishing scattering amplitudes of charge-carrying
particles by the recipe for photon-inclusive cross sections.

A more radical behavior is expected in the massless limit of
self-interacting vector mesons. The conjectured confinement means that only
correlations which contain solely point-local hadron and gluonium composites
as well as "string-bridged" $q-\bar{q}$ pairs will survive this limit
whereas correlation containing string-local gluon or quark fields vanish.
From the analogy with the vanishing QED scattering amplitudes for
electrically charged particles with a finite number of participating photons
one expects that there will be logarithmic infrared divergencies in the
presence of gluon or quark operators whose summation of leading perturbative
contributions for $m<<1$ will lead to functions which vanish in the massless
limit.

But does this conjecture have a chance to be confirmed by perturbative
calculations? Certainly not in GT, since off-shell correlation functions in
covariant gauges are known to remain infrared finite \cite{Hollands}. But
the directional fluctuations of string-local fields have stronger infrared
ramifications than the point-local fields in covariant gauges of GT. They
are expected to be especially strong in the presence of self-interacting
vector potentials. The lowest order candidate for testing the possible
presence of an off-shell logarithmic divergence for $m\rightarrow 0~$would
be the $e$-dependent second order $F$-$F~$propagator. The off-shell
perturbation theory for string-local fields is presently under construction.

\section{Gauge theory and local quantum physics}

Despite significant conceptual differences between GT and SLFT there are
also formal analogies. A comparison between these two ways of describing QFT
of $s=1$ fields leads to a better understanding of their mutual relation. In
the absence of interactions the differences only affects a fine point in the
interpretation of causality in gauge invariant Wilson loops (section 3), but
they increase in the presence of interactions. The construction of the
globally gauge invariant S-matrix is a good illustration of the conceptual
differences between the two settings.

The CGI BRST operator formalism handles this problem in the following way.
Instead of (\ref{lin}) one writes 
\begin{equation}
A_{\mu }^{K}(x)-\partial _{\mu }\phi ^{K}=:A_{\mu }^{P,K}(x)
\end{equation}%
where $K~$refers to Krein space, $A_{\mu }^{K}$ is the massive
vector-potential in the Feynman gauge, $\phi ^{K}$ is a massive free scalar
field with the opposite sign in its two-point function (the auxiliary "St%
\"{u}ckelberg field") and $A_{\mu }^{P,K}$ is a substitute for the Proca
potential which does not explicitly appear in the Krein Fock but plays the
role of creating Wigner particle states in expectation values or
matrix-elements of gauge invariant operators between such states to space.
Whereas $\phi ^{K}$ adds unphysical degrees of freedom, the SLFT sl escort $%
\phi ~$results from a rearrangement of physical degrees of freedom\footnote{%
Remember the analogy to Cooper pairs from rearrangements of condensed matter
degrees of freedom which causes the short range nature of vector potentials
in the superconducting phase.} and disappears in the massless correlations
(no $\phi $ in the reconstructed Hilbert space)..

The $u^{K}$ and $\hat{u}^{K}$ in (\ref{Krein}) are added "ghost" degree of
freedom which serve to formalize to incorporate the unitarity arguments of
't Hooft and Veltman into an operational formalism but, together with $\phi
^{K}$ the remain without physical interpretation.

As noticed by Mund (private communication), the would-be Krein space Proca
field has indeed the same two-point function as its Hilbert space
counterpart; but being an object in a larger Krein space (the tensor product
of two Krein spaces for $s=1$ $~$with $s=0$ extended by the ghosts) it
"leaks" into the tensor-product Krein space, i.e. its "one-particle vector" $%
\left\vert p\right\rangle ^{K}~$as a vector in Krein space is not easily
recognizable as a Krein space description of a Wigner particle.

Using the ghost rules (\ref{Krein}) one defines a $L^{K},Q_{\mu }^{K}$ pair
which fulfills \cite{Scharf} 
\begin{eqnarray}
&&\mathfrak{s}L^{K}-\partial ^{\mu }Q_{\mu }^{K}=0,\text{ }hence~\mathfrak{s}%
S^{(1)}=0  \label{gauge} \\
&&\mathfrak{s}TL^{K}L^{K\prime }-\partial ^{\mu }TQ_{\mu }^{K}L^{\prime
K}-\partial ^{\prime \mu }TL^{K}Q_{\mu }^{\prime K}=0,~hence~\mathfrak{s}%
S^{(2)}=0  \notag
\end{eqnarray}%
In analogy to the $e$-independence (\ref{Q}), but with the significant
conceptual difference that, whereas $d~$\textit{acts on the individual
spacetime string directions}, the globally acting BRST nilpotent $\mathfrak{s%
}~$has has no physical interpretation of its own, but serves to extract a
physical S-matrix from an unphysical point-local Krein space description.

The formal analogy of the SLFT $Q~$formalism (\ref{Q}) with (\ref{gauge})
stands in contrast to the quite different behavior of the two $Q^{\prime }s$
in the massless limit which only exists for $Q^{K}$. The gauge dependent
zero mass matter is fictitious, there is no interacting physical pl electron
field. The pl matter field which can still be defined as a very singular
object in the sl setting disappears in the massless limit. Whereas the gauge
formalism provides no warning concerning fundamental conceptual changes (the
breakdown of the Wigner-Fock particle setting and scattering theory), the sl
setting clearly signals this through the breakdown of the $Q$-formalism and
the message that the only object which may survive the infrared limit are
the correlation functions. This is a reminder that the that the only
physical (positivy preserving) zero mass free field objects are the massless
limits of the free sl massive correlation functions.

Since there is no presently known way to formulate collision theory in the
presence of infrared problems, the only option is to use the Bloch-Nordsiek
presriptions in the adaptation to QED in \cite{YFS} which results in cross
sections which depende on the photon energy resolution of the photon
counter. A spacetime understanding of "infra-particles" and their collisions
in QED within the principles of QFT (i.e. in terms of sl fields in Hilbert
space) is still missing.

Whereas the application of the SLFT Hilbert space formalism is still in its
initial stages, the CGI adaptation of BRST gauge theory already exists since
the early 90s \cite{Scharf} (see \cite{BDSV} for a recent account). Its main
purpose has been a clarification of the Higgs issue. That formulation of the
BRST GT is formally close to the present Hilbert space setting. An attempt
in \cite{Du-S} to use it as a platform for Hilbert space formulation of
Wigner particles turned out to be premature; the central role of the clash
between point-like localization and Hilbert space positivity and its
resolution in terms of string-local fields was only recognized afterwards in 
\cite{MSY} \cite{P-Y} \cite{Rio} and more recently in \cite{vector}.

The formal similarities between CGI and SLFT end when it comes to the
construction of the $e$-independent interaction densities $(TL..L)^{P}$ in
terms of their string-local counterparts (\ref{indep}). Together with (\ref%
{psi}) they enter in the construction of off-shell vacuum expectation values
of string-local fields and their zero mass limits. This is still virgin
territory within the reach of SLFT but outside the physical range of GT.

There are some puzzling observation which came from the CGI perturbative
implementation of on-shell gauge invariance $\mathfrak{s}S=0.$ Quite
unexpectedly one finds that if one start with the most general ansatz of an
interaction density of mutually self-interacting massive vector mesons of
the same mass (omitting the superscript $K$)%
\begin{equation}
L=\sum f_{abc}F^{a.\mu \nu }A_{\mu }^{b}A_{\nu }^{c}  \label{f}
\end{equation}%
with arbitrary real coupling coefficient $f_{abc}$, the implementation of
operator gauge invariance (\ref{gauge}) up to second order requires these
coefficients to satisfy the Lie-algebra relation of the adjoint
representation. This continues to hold if one allows the masses to be
different and uses the most general first order couplings within the
power-counting restriction, including the $H$-coupling which one needs for
the preservation of second order renormalizability. One obtains the same
relations between couplings and masses as those obtained by imposing the
formal SSB prescription on the massless gauge invariant Lagrangian \cite%
{BDSV}, but this time from merely imposing the BRST gauge invariance (\ref%
{gauge}) on the tree approximation up to third order.

At first sight I found this very impressive, but in phone conservations and
email exchanges Raymond Stora (who sadly passed away in June this year)
convinced me that one should not be surprised but rather view this as the
imprint of classical gauge symmetry (connections,fibre bundles) on the BRST
formalism. The real challenge, according to Stora, is the derivation of this
Lie algebra structure from first principles of QFT.

In the new SLFT setting which is based on only first principles it is easy
to formally derive the Lie-algebraic structure in the zero mass limit \cite%
{vector}. But since the Hilbert space $Q^{\prime }s~$diverge in the massless
limit the result is at best a consistency argument. Only a derivation in the
presence of self-interacting massive vector mesons can settle this problem
(the computations have not been finished), but the formal analogies between
the two methods leaves little doubts about its validity.

A Lie-algebra structure of mutual couplings between self-interacting vector
mesons from first principles would mean a symmetry relation between coupling
strengths which is not imposed but rather the result of the spacetime
quantum causality properties of QFT without any support through quantization
from classical gauge theory. This is very different from our usual
understanding of inner symmetries according to which every symmetric theory
can be converted into a less symmetric model (more coupling parameters)
while maintaining its field content.

At this point it is helpful to recall the standard conceptual relation
between the causal localization principles underlying QFT and internal
symmetries. If one starts from a QFT in which local fields transforms under
a unitary compact group which leaves the vacuum state invariant, the
observable algebra associated with the field algebra is simply the fixed
point algebra of the field algebra under the action of the symmetry group.

The important conceptual enrichment through the Doplicher-Haag-Roberts
superselection theory was that starting from an observable algebra with a
causal localization structure (a "net of spacetime localized algebras" \cite%
{Haag}) one can reconstruct a "field algebra" (with a nontrivial compact
group acting on it) from the superselection structure of its inequivalent
unitary representations. Observable algebras are defined solely in terms of
spacetime causality concepts of operator algebras; no other group than the
spacetime Poincar\'{e} transformations play a role in their definition. Yet
an observable algebra leads in a rather canonical way to a field algebra
together with a compact group acting on it.

This construction is somewhat astonishing since at first glance the causal
localization properties of local observables seem to have no connection with
symmetry groups; the net of localized observable algebras is covariant under
the Poincar\'{e} group (including the TCP operation) but there is no direct
indication of a symmetry of the full QFT associated to these observables.
Yet the classification of unitarily inequivalent local representations of
the spacetime-indexed set of local algebras (the local superselection
sectors of the observable algebra) leads to a field algebra and a compact
groups acting on it whose structure is fully determined by the seemingly
structureless local observables..

There is no analog in classical physics; all the classical Lagrangians with
internal symmetries are obtained by reading back QFT into the classic realm;
classical Maxwell theory and Einstein's field theory of gravitation do not
contain internal symmetries. The quantum concept of inner symmetry entered
particle physics through Heisenberg's introduction of isospin into nuclear
physics.

In a perturbative view of QFT with inner symmetries one can always think
about keeping the same field content but using a less symmetric coupling
(more independent coupling parameters). In fact the terminology "symmetry"
is only meaningful in contrast with less or no symmetry. The apparent high
symmetry of self-interacting $s=1$ vector potentials has no analog for
interactions between $s<1$ fields.

The construction uses only the spacetime causal localization properties; the
group theory is hidden in the composition structure ("fusion rules") of the
localization-preserving inequivalent representations (endomorphisms) of the
observable algebra \cite{Haag}. In this way the origin of global symmetries
(inner symmetries) is fully accounted for in terms of spacetime quantum
localization concepts.

This is very different for local gauge symmetries. Using our insight into
the relation between localization and Hilbert space positivity for
interactions involving $s\geq 1~$interactions we may say local gauge theory
results from forcing a $s=1~$Hilbert space setting which requires to use
string-local fields, into a point-local setting. The prize for this is that
the pointlike fields act in a Krein space and that we can extract physical
objects by noticing that the pointlike formalism comes with an unphysical
but nevertheless useful symmetry which acts in Krein space and whose fixed
point algebra is the physical algebra of \textit{local observables} which is
generated by point-local physical fields. Although the SLFT Hilbert space
setting has no gauge symmetry, it contains a Lie-algebraic structure in its $%
f_{abc}$ couplings between string-local fields.

This explains why theoreticians using methods of algebraic QFT had \cite%
{Haag} problems with gauge symmetries and why a foundational understanding
requires new ideas \cite{Bu-Ro}. The hope is that perturbative observation
in a Hilbert space setting can be combined with nonperturbative algebraic
results from LQP in order to obtain further clarification. It would be very
interesting to see whether the perturbative results in this paper can be
backed up by structural theorems about possible connection between
interacting QFTs involving $s=1$ fields and the need to go beyond the
point-local setting of Wightman fields.

\section{An Outlook}

As stated in the introduction, the principle motivation for writing this
paper is to direct attention to the beginnings of a new development in QFT
whose aim is to preserve renormalizability within a Hilbert space setting
for interactions involving $s=1$ and possibly higher spin fields. In the
present attempt we omitted mathematical details and focused on those
conceptual properties which distinguish the new setting from that of gauge
theory.

The clash between zero mass point-local vector potentials and Hilbert space
positivity, which is resolved by the use of covariant string-local
potentials, suggests to use such potentials also in the massive case. In
this way one constructs string-local siblings of the point-local Proca
potentials which pass smoothly to their massless counterpart. The important
gain of such a construction is that the short distance behavior is lowered
from $d_{sd}^{P}=2$ to $d_{sd}=1,~$a reduction which gauge theory achieves
at the prize of abandoning Hilbert space positivity and working instead in a
Krein space.

The use of this observation leads to an extension of renormalization theory
to interacting string-local vector mesons. Since $d_{sd}=1$ for all sl $%
s\geq 1,$ the mere fulfillment of the $d_{sd}^{int}\leq 4~$power counting
restriction for controlling short distances is easy, but the danger is now
that the interaction leads to \textit{completely delocalized} fields in
higher orders. In order to avoid this the first order string-local
interaction density $L$ must be part of a so-called $L,V_{\mu }~$pair which
turns out to be a strong restriction. Its fulfillment in every order leads
to normalization conditions which must be fulfilled in all orders of $s=1$
interactions. This has a formal similarity with the fulfillment of BRST
invariance $\mathfrak{s}S=0$ in every order of perturbation theory.

The new setting does not invalidate gauge theory but it highlights its
restricted physical range. GT permits a correct description of the S-matrix
of interaction massive vector mesons and local observables, but the physics
behind gauge-variant fields remain outside its physical range. This includes
the problem of quark confinement and also the spacetime understanding of
scattering problems in QED. Within the confines of gauge invariance, gauge
theory is a very successful placeholder of a QFT of $s=1~$particles.

Disregarding structural theorems (TCP, Spin\&Statistics, derivation of large
time scattering,...) whose proof requires Hilbert space positivity, the BRST
perturbative formulation accounts for the perturbative gauge-invariant local
observables (field strength, currents) and, what guarantied the success of
GT for the Standard Model, the perturbative unitary S-matrix. What is
missing are the matter fields which relate the world of the causal
localization principles of QFT with the measurable world of Wigner
particles. Physical consequences of long distance behavior (infrared
problems) are outside the physical range of GT and must be investigated in
the Hilbert space setting of sl fields. Whereas one expects that the
asymptotic freedom property of QCD in the gauge setting will be confirmed in
the sl setting, the long distance behavior remains outside GT's physical
range.

In order to go beyond gauge theory the SLFT Hilbert space setting for the
S-matrix must be extended to the calculation of string-local correlation
functions, a task which goes significantly beyond the construction of the
string-independent S-matrix. Here consistency demands that the counterpart
of the on-shell string independence should be the independence of vacuum
expectation values of string-local fields on $e^{\prime }s$ of \textit{inner}
propagator lines. The perturbative formulation of this idea is presently
being studied.

\begin{acknowledgement}
This work is part of an ongoing project with Jens Mund; its main purpose is
to strengthen historical roots and interconnections with other results. I am
deeply indebted for his advice and suggestions. For informations about the
early history of the standard model I owe thanks to Jos\'{e} Gracia-Bondia
and Joseph V\'{a}rilly. Raymond Stora, who sadly passed away on June 20th
this year (2015), has been my mentor during the last two years; his last
mail (always surface mail) is dated from 30.05. 2015.
\end{acknowledgement}

\end{document}